\documentclass[final,sort&compress,3p,times,UTF8]{elsarticle}
\usepackage{amssymb}
\usepackage{amsmath}
\usepackage{graphicx}
\usepackage{dcolumn}
\usepackage{bm}
\usepackage{subfigure}
\usepackage{float}
\usepackage{multirow}
\usepackage{multicol}

\journal{XXX}

\begin{document}
\begin{frontmatter}

\title{
Compound WKI-SP hierarchy and multiloop soliton solutions}

\author[]{Xiaorui Hu}

\author[]{Tianle Xu}

\author[]{Junyang Zhang}

\author[]{Shoufeng Shen\corref{cor1}}
\ead{}

\address[label1]{Department of Applied Mathematics, Zhejiang University of Technology, Hangzhou 310023, Zhejiang, China}

\cortext[cor1]{mathssf@zjut.edu.cn}

\begin{abstract}

The generalized hierarchies of compound WKI-SP
(Wadati-Konno-Ichikawa and short pulse) equations are presented. The proposed integrable nonlinear equations include the WKI-type equations,
the SP-type equations and the compound generalized WKI-SP equations. A chain of hodograph transformations are established
to relate the compound WKI-SP equations with the MKdV-SG (modified Korteweg-de Vries and sine-Gordon) equations. As
applications, the multiloop soliton solutions of one compound WKI-SP (named WKI-SP$^{(1,1)}$) equation
are obtained. We emphasize on showing abundant solitonic behaviors of two loop solitons.
The role of each parameter plays in the movement of two-loop solion are shown detailedly
in a table.

\end{abstract}

\begin{keyword}

WKI-SP equation,\ MKdV-SG equation,\ hodograph transformation,\ loop soliton.\\

\end{keyword}

\end{frontmatter}

\section{Introduction}

A remarkable development in our understanding of a
certain class of nonlinear integrable equations
known as the consistency condition for a system of linear differential
equations has taken place in the past
decade. This zero curvature equations on real Lie algebras such as sl(2,$\mathbb{R}$)and so(3,$\mathbb{R}$) lay the foundation for constructing soliton hierarchies. The most famous one is the Ablowitz-Kaup-Newell-Segur (AKNS) hierarchy, including the fundamental Korteweg-de Vries (KdV) equation, the the modified KdV (MKdV) equation, the sine-Gordon (SG) equation, the nonlinear
Schr\"{o}dinger (NLS) equation and so on.

In 1979, Wadati, Konno and Ichikawa (WKI) \cite{WKI79-1,WKI79-2}  constructed a new series of integrable nonlinear evolution equations, such as
\begin{eqnarray}
u_t+\bigg[\frac{u_x}{(1+u^2)^{3/2}}\bigg]_{xx}=0. \label{1-1}
\end{eqnarray}
The nonlinear terms of Eq.\eqref{1-1} have saturation effects. Eq.\eqref{1-1}
describes nonlinear transverse oscillations of elastic
beams under tension \cite{YKM1981}.
In the many past decades, lots of research have been done for
this WKI equation \eqref{1-1}. For example, an infinite number of conservation laws and the Hamiltonian form was found by  Wadati et al \cite{WKI79-1}.
The one-soliton and two-soliton solutions were obtained by using inverse scattering transform method in Ref.\cite{Konno1984}. Qu and Zhang \cite{Qu-WKI-2005} derived the WKI
model from motion of curves in E$^3$
and gave the corresponding group-invariant solutions. The Darboux transformation is
proposed by Zhang et. al. \cite{ZYS-2017} and the direct scattering problem with box-like initial
value was considered by Tu and Xu \cite{Xu-WKI-2021}.

Recently, the systems of short pulse type have attracted considerable attention due to their important applications in physics. The very first such equation may be the short pulse (SP) equation
\begin{eqnarray}
u_{xt}=u+\frac{1}{6}(u^3)_{xx}, \label{1-2}
\end{eqnarray}
which possesses a Lax pair of the WKI type. The SP equation \eqref{1-2} is proposed by Sch\"{a}fer and Wayne \cite{SP-2004,SP-2005} as an alternative (to the NLS equation)
model for approximating the evolution of ultrashort intense infrared pulses in silica
optics. It is shown in \cite{SP-2005} by numerical simulations that the SP equation can be
successfully used for describing pulses with broad spectrum.
It turns out that the SP equation made its first appearance in Rabelo's paper \cite{SP-1989}
in his study pseudospherical surfaces. The SP and two-component SP (real or complex) equations  are proposed as special integrable cases in the negative WKI hierarchy for the first
time in Refs.\cite{qiao1, qiao2}. Multi-soliton solutions and the Cauchy
problem for a two-component SP system are given in Ref.\cite{qiao3}.
The SP equation is proved integrable \cite{Sakovich-2005,Brunelli-2006} since it admits Lax pair, recursion operator, and bi-Hamiltonian structure and various exact solutions. Different approaches have been employed to construct solutions for the SP equation.
In particular, Anton Sakovich and Sergei Sakovich \cite{Sakovich-2006} found an exact nonsingular solitary wave solution from the breather solution of the SG equation by means of a transformation between these two integrable equations. Kuetche et al \cite{Kuetche-2007} calculated the
two-loop soliton solution by use of Hirota bilinear method and Hodnett-Moloney approach.
Matsuno \cite{Matsuno-2007,Matsuno-2008} obtained the multiloop soliton, multibreather and the periodic solutions. Liu et al \cite{He-2017} constructed the $N$-fold Darboux transformation given by determinants. The Riemann-Hilbert approach \cite{Monvel-2017,XuJ-2018} was applied to the SP equation and the long-time behavior of the solution was studied. Liu and Mao \cite{LiuQP-2020} established the $N$-B\"{a}cklund transformation and worked out various solutions including loop solitons, breather solutions, and their interaction solutions.
Feng et al \cite{FengBF-2010,FengBF-2014} studied its integrable discretizations and
both semidiscrete and full-discrete systems were obtained.

In this paper, we focus on the construction of the generalized hierarchies of compound
WKI-SP equations from the consistency condition of two linear differential system.
It is known that both the WKI equation and the SP equation admit loop soliton solutions
\cite{Konno1984,Kimiaki-1983}.
It is noted that they can be transformed to certain soliton equations which admit smooth soliton solutions through hodograph transformations\cite{FengBF-2011}. For example, the WKI elastic beam equation is transformed to the potential mKdV equation.  The SP equation can be transformed into the integrable SG equation. Hence we want to seek for a connection between the compound MKdV-SG equations (given by Gu in Ref.\cite{Gu-1,Gu-2}) and the WKI-SP equations. As applications, we will calculate multiloop soliton solutions for one  WKI-SP equation.

\section{Compound equations of the WKI-SP type}

Many of the hardest problems and most interesting phenomena being studied by
 mathematicians, engineers and physicists are nonlinear in nature. Often, these
 phenomena can be modeled (and there is good reason to believe that these models
 are accurate) by nonlinear partial differential equations and to be sure, it will be
 many years to come before we have the mathematical sophistication to handle
 these equations completely. In the last decades, the inverse scattering transform (IST)
 has been employed to solve many physically significant equations. Due to
 the similarity of the method itself to Fourier transforms, this theory can be considered a
 natural extension of Fourier analysis to nonlinear problems. Along with the IST,
 a systematic method has developed which allows one to identify certain important
 classes of evolution equations which can be solved by the method of inverse
 scattering. These evolution equations are expressed as the consistency condition
 \begin{eqnarray}
 U_t-V_x+[U,V]=0,  \quad [U,V]=UV-V U          \label{eq1}
 \end{eqnarray}
 for a system of linear differential equations
 \begin{eqnarray} \label{eq2}
 	\Phi_x=U(x,t;\lambda)\Phi,  \quad \quad
     \Phi_t=V(x,t;\lambda)\Phi,
 \end{eqnarray}
 where $U$, $V$ and $\Phi$ are complex $N \times N$ matrixes. The matrixes
 $U$ and $V$ are usually rational functions of the parameter $\lambda$.

\subsection{AKNS-type and WKI-type Lax pair}

To increase the readability of the article, we firstly recall the similar results for
the known AKNS hierarchy and the WKI hierarchy.

In 1973, a wide class of nonlinear evolution equations were presented by Ablowitz, Kaup,Newell and Segur (AKNS) \cite{AKNS1}, where $U$ and $V$ read

 \begin{eqnarray} \label{eq3}
 U=\left(\begin{array} {cc}
 		-i \lambda & q(x,t)  \cr
 		r(x,t) & i \lambda
 \end{array}\right), \quad
 V=\left(\begin{array} {cc}
 		A(x,t;\lambda) &  B(x,t;\lambda)  \cr
 		C(x,t;\lambda) & -A(x,t;\lambda)
 \end{array}\right).
 \end{eqnarray}

The corresponding consistency condition \eqref{eq1}  leads to the flowing equations
\begin{eqnarray}  \label{eq4}
&& A_x=q c-r B,\\
&&q_t-B_x-2i\lambda B-2 q A=0,\\
&&r_t-C_x+2i\lambda C+2rA=0.
\end{eqnarray}

For example, AKNS set
\begin{eqnarray} \label{eq5}
A=\sum _{j=0} ^{N}a_j \lambda^{N}, \quad
B=\sum _{j=0} ^{N}b_j \lambda^{N}, \quad
C=\sum _{j=0} ^{N}c_j \lambda^{N}.
\end{eqnarray}
To illustrate, take $N=3$. By solving Eqs.\eqref{eq4}, AKNS found
\begin{eqnarray}
&&A=a_3\lambda^3+a_2\lambda^2+(\frac{1}{2}a_3qr+a_1)\lambda+  \frac{1}{2}a_2qr
-\frac{i}{4} a_3(q r_x-q_x r)+a_0,   \\
&&B=i a_3q\lambda^2+(ia_2q-\frac{1}{2}a_3q_x)\lambda+(ia_1q+\frac{i}{2}a_3q^2r
-\frac{1}{2}a_2q_x-\frac{i}{4}a_3q_{xx}),   \\
&&C= i a_3r\lambda^2+(ia_2r+\frac{1}{2}a_3r_x)\lambda+(ia_1r+\frac{i}{2}a_3qr^2
+\frac{1}{2}a_2r_x-\frac{i}{4}a_3r_{xx})    \label{eq6}
\end{eqnarray}
together with the evolution equations
\begin{eqnarray}
&&q_t+\frac{i}{4}a_3(r_{xxx}-6qrq_x)+\frac{1}{2} a_2(q_{xx}-2q^2r)-ia_1q_x-2a_0q=0,\\
&&r_t+\frac{i}{4}a_3(r_{xxx}-6qrr_x)-\frac{1}{2} a_2(r_{xx}-2qr^2)-ia_1r_x+2a_0r=0.
\end{eqnarray}

\emph{\textbf{Case I.}} As a special case, let $r=-q$  and
\begin{eqnarray}
a_0=a_1=a_2=0, \quad a_3=-4i.
\end{eqnarray}
Then AKNS got the modified mKdV equation
\begin{eqnarray}
 q_t+6q^2q_x+q_{xxx}=0.
\end{eqnarray}

\emph{\textbf{Case II.}} In the same way, by taking
\begin{eqnarray}
A=\frac{a(x,t)}{\lambda},\quad
B=\frac{b(x,t)}{\lambda},
C=\frac{c(x,t)}{\lambda},
\end{eqnarray}
AKNS found
\begin{eqnarray}
q_{xt}=-4iaq,\quad r_{xt}=-4iar,\quad
a_x=\frac{i}{2}(qr)_t.
\end{eqnarray}
As a special but important case, we list the following SG equation
 \begin{eqnarray}
u_{xt}=\sin u
\end{eqnarray}
 with
\begin{eqnarray}
a=\frac{i}{4} \cos u, \quad b=c=\frac{i}{4}\sin u, \quad r=-q=\frac{1}{2}u_x.
\end{eqnarray}

In 1979, Wadati, Konno and Ichikawa \cite{WKI79-1,WKI79-2} proposed a generalization of the inverse scattering formalism (especially $U$) and found a new series of
integrable nonlinear  evolution equations. In fact, they considered
 \begin{eqnarray}
 U=\left(\begin{array} {cc}
 		-i \lambda & \lambda q  \cr
 \lambda r & i \lambda
 \end{array}\right), \quad
 V=\left(\begin{array} {cc}
 		A &  B  \cr
 		C & -A
 \end{array}\right).
 \end{eqnarray}

The corresponding consistency condition \eqref{eq1}  leads to the flowing equations
\begin{eqnarray}
&& A_x=\lambda (q C-r B),\label{eq22}  \\
&&\lambda q_t-B_x-2i\lambda B-2\lambda q A=0,\label{eq23}  \\
&&\lambda r_t-C_x+2i\lambda C+2\lambda rA=0. \label{eq24}
\end{eqnarray}

\emph{\textbf{Case III.}} As a special case, by choosing
\begin{eqnarray}
&& A=-\frac{4i}{\sqrt{1-rq}}\lambda^3+\frac{rq_x-qr_x}{(1-rq)^{3/2}} \lambda^2, \\
&& B= \frac{4q}{\sqrt{1-rq}}\lambda^3+\frac{2iq_x}{(1-rq)^{3/2}} \lambda^2
-\bigg(\frac{q_x}{(1-rq)^{3/2}} \bigg)_x\lambda,\\
&&C= \frac{4r}{\sqrt{1-rq}}\lambda^3-\frac{2ir_x}{(1-rq)^{3/2}} \lambda^2
  -\bigg(\frac{r_x}{(1-rq)^{3/2}}\bigg)_x\lambda,
\end{eqnarray}
they obtained
\begin{eqnarray}
&&q_t+\bigg(\frac{q_x}{(1-rq)^{3/2}}\bigg)_{xx}=0,\\
&&r_t+\bigg(\frac{r_x}{(1-rq)^{3/2}}\bigg)_{xx}=0.
\end{eqnarray}
For $r=-q$, above equation is reduced to the typical WKI equation \eqref{1-1}.

\emph{\textbf{Case IV.}} Moreover, as another special case, by choosing
$r=q=iu_x$ and
\begin{eqnarray}
A= \frac{i\lambda}{2}u^2+\frac{1}{4i\lambda},\quad
B=\frac{i\lambda}{2}u^2u_x-\frac{1}{2}u,\quad C=\frac{i\lambda}{2}u^2u_x+\frac{1}{2}u,
\end{eqnarray}
one can get the known SP equation \eqref{1-2}. This Lax pair for \eqref{1-2} was firstly
discovered by Sakovich A and Sakovich S \cite{Sakovich-2005} in 2005.

\subsection{WKI-SP type integrable equations}

In this paper, we focus on a series of compound equations of the WKI-SP type, which will
be derived on the basis of the linear system \eqref{eq2}.
For convenience, we rewrite $U$ as
\begin{eqnarray}
U=\left(\begin{array} {cc}
		\lambda & \lambda p(x,t)  \cr
		\lambda p(x,t) & -\lambda
\end{array}\right).
\end{eqnarray}
Eqs.\eqref{eq22}-\eqref{eq24} become
\begin{eqnarray}
&& A_x=\lambda p (C-B),\\
&&\lambda p_t-B_x+2\lambda B-2\lambda p A=0,\\
&&\lambda p_t-C_x-2\lambda C+2\lambda p A=0.
\end{eqnarray}

Consequently, we have
\begin{eqnarray}
&&C-B=\frac{A_x}{\lambda p}, \label{2-6}\\
&&C+B=-\frac{(C-B)_x}{2\lambda}+2pA,\label{2-7} \\
&&p_t=\frac{(C+B)_x}{2\lambda}+(C-B).\label{2-8}
\end{eqnarray}

Different from the previous references, we select $A$ as follows:
\begin{eqnarray}
	A=\sum _{j=0} ^{n+m}a_j \lambda^{2n-2j+1}, \quad \quad (n\geq0, m\geq0).\label{2-2}
\end{eqnarray}

Substituting \eqref{2-6} and \eqref{2-7} into \eqref{2-8} and then expanding it into a power series of $\lambda$, we obtain
\begin{eqnarray}
p_t=-\frac{1}{4}\bigg(\frac{a_{n-1,x}}{p}\bigg)_{xx}+(pa_n)_x+\frac{a_{nx}}{p} \label{10}
\end{eqnarray}
together with the following two sets of recursive formulas
\begin{eqnarray}
&&(pa_0)_x+\frac{a_{0x}}{p}=0, \label{2-10}\\
&&(pa_j)_x+\frac{a_{j,x}}{p}=\frac{1}{4}\bigg(\frac{a_{j-1,x}}{p}\bigg)_{xx}, \quad (j=1,2,\ldots,n-2,n-1), \label{2-11}
\end{eqnarray}
and
\begin{eqnarray}
&&-\frac{1}{4}\bigg(\frac{a_{n+m,x}}{p}\bigg)_{xx}=0, \label{2-12}\\
&&\frac{1}{4}\bigg(\frac{a_{j,x}}{p}\bigg)_{xx}=(pa_{j+1})_x+\frac{a_{j+1,x}}{p}, \quad (j=n+m-1,\ldots,n+1,n).\label{2-13}
\end{eqnarray}

It is seen that $\{a_0,a_1, \ldots, a_{n-1}\}$ and $\{a_n,a_{n+1}, \ldots, a_{n+m-1},a_{n+m}\}$ are determined by Eqs.\eqref{2-10}-\eqref{2-11} and Eqs.\eqref{2-12}-\eqref{2-13}, respectively.
For convenience, we denote the defined equation in \eqref{10} as WKI-SP$^{(n,m)}$ equation for the concrete $n$ and $m$.

Furthermore, by solving Eqs.\eqref{2-10}-\eqref{2-11}, we obtain
\begin{eqnarray}
&&a_0=\frac{a_{0}^{0}(t)}{\sqrt{1+p^{2}}},\label{15}\\
&&a_j=\frac{1}{4\sqrt{1+p^2}}	\partial^{-1}\bigg(\frac{p}{\sqrt{1+p^2}} \bigg(\frac{a_{j-1,x}}{p}\bigg)_{xx}\bigg), \quad (j=1,2,\ldots,n-1), \label{16}
\end{eqnarray}
which imply
\begin{eqnarray}
&&a_1=-\frac{a_0^0(t)\biggl[(p^3+p)p_{xx}-\frac{1}{2}(6p^2+1)p_x^2\biggr]}
{4(1+p^2)^{\frac{7}{2}}},\label{17}
\end{eqnarray}
\begin{equation}	
	\begin{split}
		 &a_2=-\frac{a_0^{0}(t)}{32(1+p^2)^\frac{13}{2}}\bigg[2p(1+p^2)^3p_{xxxx}-2(1+p^2)^2(15p^2+1)
p_xp_{xxx}-(1+p^2)^2(20p^2-1)p_{xx}^{2}\\
		&+21p(1+p^2)(10p^2-1)p_x^2p_{xx}-\frac{21}{4}(40p^4-16p^2-1)p_x^4\bigg],\label{18} \\
& \quad \vdots \nonumber
	\end{split}
\end{equation}

By solving Eqs.\eqref{2-12}-\eqref{2-13}, we have
\begin{eqnarray}
&&a_{n+m}=a_{n+m}^0(t), \label{21}\\
&&a_{j}=4\partial^{-1}\bigg(p\partial^{-2}\bigg(\frac{a_{j+1,x}}{p}\bigg)\bigg)
+4\partial^{-1}(p\partial^{-1}(pa_{j+1})), \label{22}\\
&& (j=n+m-1,n+m-2,\ldots,n), \nonumber
\end{eqnarray}
which
imply
\begin{eqnarray}
&&a_{n+m}=a_{n+m}^0(t),\\
&&a_{n+m-1}=2a_{n+m}^0(t)(\partial^{-1}p)^2,\\
&&a_{n+m-2}=a_{n+m}^0(t)\bigg[\frac{2}{3}(\partial^{-1}p)^4+16(\partial^{-1}p)(\partial^{-3}p)-8(\partial^{-2}p)^2\bigg],\\
&&\begin{split} &a_{n+m-3}=a_{n+m}^0(t)\bigg\{\frac{4}{45}(\partial^{-1}p)^6+\frac{32}{3}\biggl[(\partial^{-1}p)^3(\partial^{-3}p)+(\partial^{-1}p)\bigg(\partial^{-2}(\partial^{-1}p)^3\bigg)-(\partial^{-2}p)\bigg(\partial^{-1}(\partial^{-1}p)^3\bigg)\\
&-\frac{3}{2}\partial^{-1}\bigg(p(\partial^{-2}p)^2\bigg)\biggr]
+64\bigg[(\partial^{-1}p)(\partial^{-5}p)-(\partial^{-2}p)(\partial^{-4}p)
+\frac{1}{2}(\partial^{-3}p)^2\bigg]\bigg\},
\end{split}\\
&&\quad \vdots  \nonumber
\end{eqnarray}
In order to express the fomulas clearly, we write $\partial^{-1}$ instead of $\int^{x} \cdot dx$ sometimes.

\subsubsection{SP-hierarchy for $n=0$}

If $n=0$, \eqref{2-2} becomes
\begin{eqnarray}
A=\sum _{j=0} ^{m}a_j \lambda^{-2j+1}=a_0\lambda+a_1\lambda^{-1}+a_2\lambda^{-3}+\cdots+
a_m\lambda^{-(2m-1)}.
\end{eqnarray}

Then we obtain the generalized hierarchy of SP equations
\begin{eqnarray}
p_t=(pa_0)_x+\frac{a_{0x}}{p}.\label{28}
\end{eqnarray}

Here $a_0$ is determined by the recursive formulas \eqref{21} and \eqref{22} with $n=0$.
Now we give some special cases of the SP hierarchy.

\textbf{\emph{Case 1.}} For $m=1$ and $p=u_{x}$, we have
\begin{eqnarray}
a_1=a^{0}_1(t),\quad a_0=2a_1^0(t)u^2,
\end{eqnarray}
which exactly give rise to the short pulse equation
\begin{eqnarray}
u_{xt}=4a_{1}^0(t)\bigg(\frac{1}{6}(u^3)_{xx}+u\bigg). \label{30}
\end{eqnarray}
According to the convention above, we label Eq.\eqref{30} as WKI-SP$^{(0,1)}$ equation.

\textbf{\emph{Case 2.}} For $m=2$ and $p=u_{xxx}$, we have
\begin{eqnarray}
a_2=a^{0}_2(t), \quad  a_1=2a_2^0(t)u^2_{xx},\quad a_0=\frac{2}{3}a_2^0(t)u_{xx}^4+16a_2^0(t)uu_{xx}-8a_2^0(t)u_x^2,
\end{eqnarray}
which yields a high-order short pulse (i.e.WKI-SP$^{(0,2)}$) equation
\begin{eqnarray}
u_{xxxt}=\frac{2}{15}a_{2}^0(t)(u_{xx}^5)_{xx}
+a_{2}^0(t)[u_{xxx}(16uu_{xx}-8u_x^2)]_{x}
+\frac{8}{3}a_{2}^0(t)u_{xx}^3+16a_{2}^0(t)u.
\end{eqnarray}
This equation is just the local form of the one given in Ref.\cite{Brunelli2005}.

\textbf{\emph{Case 3.}} For $m=3$ and $p=u_{xxxxx}$, we have
\begin{equation}
	a_3=a^{0}_3(t), \quad  a_2=2a_3^0(t)u_{xxxx}^2, \quad  a_1=\frac{2}{3}a_3^0(t)u_{xxxx}^4+16a_3^0(t)u_{xx}u_{xxxx}-8a_3^0(t)u_{xxx}^2,
\end{equation}
\begin{equation}
	\begin{split}
		 a_0=&\frac{4}{45}a_3^0(t)u_{xxxx}^6+32a_3^0(t)\bigg[\frac{1}{3}u_{xx}u_{xxxx}^3-\frac{1}{2}u_{xxx}^2u_{xxxx}^2+2uu_{xxxx}-2u_xu_{xxx}+u_{xx}^2+u_{xxxx}\partial^{-1} (u_{xxxx}^2u_{xxx})\\
		&+\frac{32}{3}u_{xxxx}\partial^{-2}u_{xxxx}^3\bigg],
	\end{split}
\end{equation}
which yields the WKI-SP$^{(0,3)}$ equation
\begin{equation}
\begin{split}
	 &u_{xxxxxt}=a_3^0(t)\bigg\{u_{xxxxxx}\bigg[\frac{4}{45}u_{xxxx}^6+32\bigg(\frac{1}{3}u_{xx}u_{xxxx}^3-\frac{1}{2}u_{xxx}^{2}u_{xxxx}^2+u_{xxxx}\partial^{-1}(u_{xxxx}^{2}u_{xxx})+\frac{32}{3}u_{xxxx}\partial^{-2}(u_{xxxx}^{3})\\
	 &+64\bigg(uu_{xxxx}-u_{x}u_{xxx}+\frac{1}{2}u_{xx}^2\bigg)\bigg)\bigg]+(u_{xxxxx}
+{u^{-1}_{xxxxx}})\bigg[\frac{8}{15}u_{xxxx}^{5}u_{xxxxx}+32u_{xxxxx}\\
	 &\bigg(u_{xx}u_{xxxx}^2-u_{xxx}^{2}u_{xxxx}+\partial^{-1}
(u_{xxxx}^{2}u_{xxx})+\frac{32}{3}u_{xxxxx}\partial^{-2}(u_{xxxx}^3)\bigg)+64uu_{xxxxx}\bigg]\bigg\}.
\end{split}
\end{equation}

\subsubsection{WKI-hierarchy for  $m=0$}

If $m=0$, \eqref{2-2} is rewritten as
\begin{eqnarray}
A=\sum _{j=0} ^{n}a_j \lambda^{2n-2j+1}=a_0\lambda^{2n+1}+a_1\lambda^{2n-1}+a_2\lambda^{-3}+\cdots+
a_n\lambda.
\end{eqnarray}

Then we obtain the generalized hierarchy of WKI equations
\begin{eqnarray}
p_t=-\frac{1}{4}\bigg(\frac{a_{n-1,x}}{p}\bigg)_{xx}+a_n^0(t)p_x, \label{35}
\end{eqnarray}
with the recursive formulas \eqref{15} and \eqref{16}.
Some special equations in the WKI hierarchy are given in the  follows.

\textbf{\emph{Case 4.}} For $n=1$ and $p=u_{x}$, Eq.\eqref{35} with \eqref{17} is exactly the
WKI equation
\begin{eqnarray}
u_{xt}=\frac{a_{0}^0(t)}{4}\left[\frac{u_{xx}}{(1+u_x^2)^\frac{3}{2}}
\right]_{xx}+a^0_1(t)u_{xx}.
\end{eqnarray}

\textbf{\emph{Case 5.}} For $n=2$ and $p=u_x$, Eq.\eqref{35} with \eqref{18} is the high-order WKI
(i.e. WKI-SP$^{(2,0)}$) equation
\begin{equation}	
	\begin{split}
	 &u_{xt}=\frac{a_0^0(t)}{16\left(1+u_x^2\right)^\frac{13}{2}}\Bigg[u_{xxxxxx}
\left(1+u_x^2\right)^{4}-20u_{x}u_{xx}u_{xxxxx}\left(1+u_{x}^2\right)^{3}-35u_{xxxx}\Bigg(u_{xxx}\left(u_x^{3}+u_x\right)-\frac{39}{7}u_{xx}^2 \\
	 &(u_x^2-\frac{1}{6})\Bigg)\left(1+u_x^2\right)^{2}+270u_{xx}\Bigg(\left(1+u_x^2\right)^2(u_x^2-\frac{1}{6})u_{xxx}^{2}-\frac{77}{18}u_{x}u_{xxx}u_{xx}^2(u_x^2-\frac{1}{2})\left(1+u_x^2\right) \\
	 &+\frac{28}{9}u_{xx}^4(u_x^4-u_x^2+\frac{1}{16})\Bigg)\Bigg]+\frac{1}{4}a_{1}^{0}\left(t\right)\left(\frac{u_{x}}{\sqrt{1+u_{x}^2}}\right)_{xxx}+a_3^0(t)u_{xx}.
\label{41}
	\end{split}
	\end{equation}

\subsubsection{Compound WKI-SP hierarchy for  $n\neq0, m\neq0$}

For $n\neq0$ and $m\neq0$, we have the generalized compound WKI-SP$^{(n,m)}$ equations \eqref{10}.
Here, we give three examples.

\textbf{\emph{Case 6.}} For $n=1, m=1$ and $p=u_{x}$, Eq.\eqref{10} gives the WKI-SP$^{(1,1)}$ equation
\begin{eqnarray}
u_{xt}=\frac{a_{0}^0(t)}{4}\left[\frac{u_{xx}}{\bigg(1+u_x^2\bigg)^\frac{3}{2}}
\right]_{xx}+a_{1}^0(t)u_{xx}+4a_{2}^0(t)\bigg(\frac{1}{6}(u^3)_{xx}+u\bigg).\label{39}
\end{eqnarray}

\textbf{\emph{Case 7.}} For $n=1,m=2$ and $p=u_{xxx}$, we obtain the following WKI-SP$^{(1,2)}$ equation
\begin{eqnarray}
u_{xxxt}=\frac{a_{0}^0(t)}{4}\left[\frac{u_{xxxx}}{(1+u_{xxx}^2)^\frac{3}{2}}
\right]_{xx}+a_{2}^0(t)\bigg(\frac{2}{15}(u_{xx}^5)_{xx}+[u_{xxx}(16uu_{xx}-8u_x^2)]_{x}
+\frac{8}{3}u_{xx}^3+16u\bigg). \label{40}
\end{eqnarray}

\textbf{\emph{Case 8.}} For $n=2,m=1$ and $p=u_x$, we obtain the following WKI-SP$^{(2,1)}$ equation

\begin{equation}	\label{WKI-SP21}
	\begin{split}
	 &u_{xt}=\frac{a_0^0(t)}{16\left(1+u_x^2\right)^\frac{13}{2}}\Bigg[u_{xxxxxx}
\left(1+u_x^2\right)^{4}-20u_{x}u_{xx}u_{xxxxx}\left(1+u_{x}^2\right)^{3}-35u_{xxxx}
\Bigg(u_{xxx}\left(u_x^{3}+u_x\right)-\frac{39}{7}u_{xx}^2 \\
	 &(u_x^2-\frac{1}{6})\Bigg)\left(1+u_x^2\right)^{2}+270u_{xx}
\Bigg(\left(1+u_x^2\right)^2(u_x^2-\frac{1}{6})u_{xxx}^{2}
-\frac{77}{18}u_{x}u_{xxx}u_{xx}^2\left(u_x^2-\frac{1}{2}\right)(1+u_x^2) \\
	 &+\frac{28}{9}u_{xx}^4(u_x^4-u_x^2+
\frac{1}{16})\Bigg)\Bigg]+\frac{1}{4}a_{1}^0 \left(t\right)\left(\frac{u_{x}}
{\sqrt{1+u_{x}^2}}\right)_{xxx}+4a_3^0(t)\left(u+\frac{1}{6}\left(u^3\right)_{xx}\right).
	\end{split}
	\end{equation}

%

\section{Hodograph transformations between MKdV-SG and WKI-SP eqautions}

In Ref.\cite{Gu-1,Gu-2}, Gu obtained the compound equations of the
MKdV-SG type
\begin{eqnarray}
q_t-\frac{1}{4}\Bigg[4a_{n}q+\bigg(\frac{a_{nx}}{q}\bigg)_x\Bigg]_x+\frac{a_{n+1,x}}{q}=0
	\label{42}
\end{eqnarray}
together with the following two sets of recursive formulas
\begin{eqnarray}
&&a_{0x}=0,\\
&&a_{hx}=\frac{1}{4}q\Bigg[4a_{h-1}q+\bigg(\frac{a_{h-1,x}}{q}\bigg)_x\Bigg]_x ,h=1,2,...\,,n  ,
\label{43}
\end{eqnarray}
and
\begin{eqnarray}
&&\Bigg[4a_{l+m}q+\bigg(\frac{a_{l+m,x}}{q}\bigg)_x\Bigg]_x=0,\\
&&\Bigg[4a_{h-1}q+\bigg(\frac{a_{h-1,x}}{q}\bigg)_x\Bigg]_x=4\frac{a_{h,x}}{q} , h=n+m,...\,,n+2 .
	\label{44}
\end{eqnarray}

In this section, we would build the hodograph transformations between some members in the above
MKdV-SG type equations and the WKI-SP$^{(n,m)}$ equation. To see it clearly, we denote
the concrete equation among \eqref{42} as MKdV-SG$^{(n,m)}$ equation for specific $n$ and $m$.

\subsection{Hodograph transformation between MKdV-SG$^{(1,1)}$ and WKI-SP$^{(1,1)}$ equation}

By taking $n=1, m=1$ and $q=-\frac{\theta_{s}}{2}$ in Eq.\eqref{42}, the author gave the  MKdV-SG$^{(1,1)}$ equation
\begin{eqnarray}
\theta_{ts}-\frac{1}{4}\alpha_0(t)\biggl(\frac{3}{2}\theta_s^2\theta _{ss}+\theta _{ssss}\biggr)-\alpha _1(t)\theta _{ss}=\alpha _2(t)\sin{\theta },
\label{3-26}
\end{eqnarray}
which is related to the motion of a nonlinear one-dimensional lattice of atoms.

Integrating \eqref{3-26} with respect to $s$ leads to
\begin{eqnarray}
\theta_t-\frac{1}{4}\alpha_0(t)\Bigg(\frac{1}{2}\theta_s^3+
\theta_{sss}\Bigg)-\alpha_1(t)\theta_s=\alpha_2(t)\int^s \sin{\theta}(s_1,t)\mathrm{d}s_1.
\label{3-27}
\end{eqnarray}

A conservation law of Eq.\eqref{3-27} is given by
\begin{eqnarray}
(\cos \theta)_{t}+\Bigg[\frac{1}{4}\alpha_0(t)\bigg(\theta_{ss}\sin{\theta}-\frac{1}{2}
\theta_s^2\cos{\theta}\bigg)-\alpha_1(t)\cos{\theta }+\frac{1}{2}\alpha_2(t)\bigg(\int^s\sin{\theta }(s_1,t)\mathrm{d}s_1\bigg)^2\Bigg]_s=0.
\label{3-28}
\end{eqnarray}

Consider the hodograph transformation
\begin{eqnarray}
x(s,t)=\int^s \cos \theta(s_1,t){\rm{d}} s_1+x_0,\quad\quad\quad\quad  t'(s,t)=t,
\label{3-29}
\end{eqnarray}
which leads to
\begin{eqnarray}
\frac{\partial}{\partial s}=\cos \theta\frac{\partial}{\partial x},\quad\quad
\frac{\partial}{\partial t}=\Bigg[\frac{1}{4}\alpha_0(t)\bigg(\frac{1}{2}\theta_s^2\cos{\theta }-\theta_{ss}\sin{\theta }\bigg)+\alpha_1(t)\cos{\theta }-\frac{1}{2}\alpha_2(t)\bigg(\int^s\sin{\theta }(s_1,t)\mathrm{d}s_1\bigg)^2\Bigg]\frac{\partial}{\partial x}+\frac{\partial}{\partial t'}.
\label{3-30}
\end{eqnarray}

Applying \eqref{3-30} to \eqref{3-28},
we obtain
\begin{eqnarray}
(\tan \theta)_{t'}-\frac{1}{4}\alpha_0(t)(\sin{\theta })_{xxx}-\frac{\alpha_2(t)}{2}\bigg(\frac{(\int^s\sin{\theta }(s_1,t)\mathrm{d}s_1)^2}{\cos{\theta }}\bigg)_x\frac{1}{\sin{\theta }}=0,
\label{3-32}
\end{eqnarray}
with
\begin{eqnarray}
\theta_s=(\cos \theta) \theta_x=(\sin\theta)_x, \quad \theta_{ss}=(\cos\theta) (\theta_s)_x=(\cos\theta) (\sin\theta)_{xx}.
\label{3-31}
\end{eqnarray}

Introduce a new dependent variable
\begin{eqnarray}
u(s,t)=\int^s \sin \theta(s_1,t){\rm{d}} s_1, \label{3-33}
\end{eqnarray}
which gives
\begin{eqnarray}
u_s=\sin\theta. \label{3-34}
\end{eqnarray}

Applying \eqref{3-30} to \eqref{3-33}, we obtain
\begin{eqnarray}
u_s=(\cos\theta)u_x. \label{3-35}
\end{eqnarray}

Combing \eqref{3-34} with \eqref{3-35} gives
\begin{eqnarray}
\tan \theta =u_x, \label{3-36}
\end{eqnarray}
then it follows
\begin{eqnarray}
\sin \theta= \frac{u_x}{\sqrt{1+u_x ^2}},\quad\quad
\cos \theta= \frac{1}{\sqrt{1+u_x ^2}}.\label{3-37}
\end{eqnarray}

Substituting \eqref{3-36} and  \eqref{3-37} into \eqref{3-32} yields
\begin{eqnarray}
u_{xt'}-\frac{1}{4}\alpha _0(t) \bigg(\frac{u_x}{\sqrt{1+u_x ^2}}\bigg)_{xxx}-\alpha _2(t)\bigg(u+\frac{1}{6}(u^3)_{xx}\bigg)=0,
\label{3-38}
\end{eqnarray}
which is just the WKI-SP$^{(1,1)}$ equation given by \eqref{39} with $a^0_1(t)=0$.

\subsection{Hodograph transformation between MKdV-SG$^{(2,1)}$ and WKI-SP$^{(2,1)}$ equation}

For $n=2,m=1$, the MKdV-SG$^{(2,1)}$ equation is given by
\begin{eqnarray}
q_{t}-\frac{1}{4}\bigg[\bigg(\frac{a_{2,s}}{q}\bigg)_{s}+4a_{2}q\bigg]_{s}+\frac{a_{3,s}}{q}
=0. \label{58}
\end{eqnarray}

By solving \eqref{43}, one obtain
\begin{eqnarray}
&&a_{0}=\alpha_{0}(t),\\
&&a_{1}=\frac{1}{2}\alpha_{0}(t)q^2+\alpha_{1}(t),	 \\ &&a_{2}=\alpha_{0}(t)\bigg(\frac{1}{4}qq_{ss}-\frac{1}{8}q_{s}^2+
\frac{3}{8}q^4\bigg)+\frac{1}{2}\alpha_{1}(t)q^2+\alpha_{2}(t).
\end{eqnarray}

Let $q=-\frac{\theta_{s}}{2}$ and choose the following special solution for $a_3$ in \eqref{44}
\begin{eqnarray}	
a_3=\frac{1}{4}\alpha_3(t)\cos \theta.
\end{eqnarray}

The MKdV-SG$^{(2,1)}$ equation \eqref{58} is expressed by

\begin{equation}
	\begin{split}
&\theta_{st}+\bigg[\frac{1}{128}\alpha_{0}(t)\bigg(-8\theta_{ssssss}-20\theta_{ss}^3-
80\theta_{s}\theta_{ss}\theta_{sss}-20\theta_{s}^{2}\theta_{ssss}-15\theta_{s}^4\theta_{ss}\bigg)
+\alpha_{1}(t)(-\frac{1}{4}\theta_{ssss}-\frac{3}{8}\theta_{s}^2\theta_{ss})-
\alpha_{2}(t)\theta_{ss}\bigg]\\
&-\alpha_3(t)\sin\theta=0.\label{63}
	\end{split}
\end{equation}

Integrating Eq.\eqref{63} with respect to $s$ leads to
\begin{eqnarray}
	 \theta_{t}-\bigg[\frac{1}{128}\alpha_{0}(t)\bigg(8\theta_{sssss}+
20\theta_{s}\theta_{ss}^2+20\theta_{s}^2\theta_{sss}+3\theta_{s}^5\bigg)
-\frac{1}{8}\alpha_{1}(t)\bigg(2\theta_{sss}+\theta_{s}^3\bigg)-\alpha_{2}(t)\theta_{s}\bigg]
=\alpha_3(t)\int^s\sin\theta(s_1,t)\mathrm{d}s_1. \label{64}
\end{eqnarray}

Multiplying both sides of Eq.\eqref{64} by $-\sin\theta$, one conservation law is given by
\begin{align}
	\begin{split}
		 &(\cos\theta)_{t}+\bigg[\frac{1}{128}\alpha_{0}(t)\bigg(12\theta_s^2\theta_{ss}\sin\theta+8\theta_{ssss}\sin\theta+4\theta_{ss}^2\cos\theta-3\theta_{s}^{4}\cos\theta-8\theta_{s}\theta_{sss}\cos\theta\bigg) \\
		 &+\frac{1}{8}\alpha_{1}(t)\bigg(2\theta_{ss}\sin\theta-\theta_s^2\cos\theta\bigg)-\alpha_{2}(t)\cos\theta+\frac{1}{2}\alpha_3(t)\bigg(\int^{s}\sin\theta(s_1,t)\mathrm{d}s_{1}\bigg)^2\bigg]_{s}=0.
	\end{split}\label{65}
\end{align}

Consider the same hodograph transformation \eqref{3-29}. It leads to
\begin{eqnarray}
	\frac{\partial}{\partial s}=\cos \theta\frac{\partial}{\partial x}, \label{66}
\end{eqnarray}
\begin{align}
	\begin{split}
		&\frac{\partial}{\partial t}=\bigg[-\frac{1}{128}\alpha_{0}(t)\bigg(12\theta_{s}^2\theta_{ss}\sin\theta+8\theta_{ssss}\sin\theta+4\theta_{ss}^2\cos\theta-3\theta_s^4\cos\theta-8\theta_{s}\theta_{sss}\cos\theta\bigg)\\
		 &\qquad-\frac{1}{8}\alpha_{1}(t)\bigg(2\theta_{ss}\sin\theta-\theta_s^2\cos\theta\bigg)+\alpha_{2}(t)\cos\theta-\frac{1}{2}\alpha_3(t)\bigg(\int^{s}\sin\theta(s_1,t)\mathrm{d}s_{1}\bigg)^2\bigg]\frac{\partial}{\partial x}+\frac{\partial}{\partial t'}.
	\end{split} \label{67}
\end{align}

We have
\begin{eqnarray}
&&	\theta_s=(\cos \theta) \theta_x=(\sin\theta)_x, \quad \theta_{ss}=\cos\theta (\theta_s)_x=\cos\theta (\sin\theta)_{xx},\\
&&	 \theta_{sss}=\cos^{2}\theta \sin\theta_{xxx}+{\cos\theta}(\cos{\theta})_{x}(\sin\theta)_{xx},\\
&&\begin{split}
	 &\theta_{ssss}=2\cos^{2}\theta(\cos\theta)_{x}(\sin\theta)_{xxx}+\cos^{3}\theta(\sin{\theta})_{xxxx}+\cos{\theta}(\cos\theta)_{x}^2(\sin\theta)_{xx}\\
	 &+\cos^{2}\theta(\cos\theta)_{xx}(\sin\theta)_{xx}+\cos^{2}\theta(\cos\theta)_{x}(\sin\theta)_{xxx}.
	\end{split}
\end{eqnarray}

Applying \eqref{66} and \eqref{67} to Eq.\eqref{65}, we obtain
\begin{eqnarray}
\begin{split}	
		 &(\tan\theta)_{t'}+\frac{\alpha_{0}(t)}{128}\bigg[24(\sin\theta)_{x}(\sin\theta)_{xx}^2+12(\sin\theta)_{x}^{2}(\sin\theta)_{xxx}+16(\sin\theta)_{xxx}(\cos\theta)_{x}^2+16(\sin\theta)_{xxx}\cos{\theta}(\cos\theta)_{xx}\\
		 &+32(\sin\theta)_{xxxx}\cos{\theta}(\cos\theta)_{x}+8(\sin\theta)_{xxxxx}\cos^{2}\theta+16(\sin\theta)_{xx}(\cos\theta)_{x}(\cos\theta)_{xx}+8(\sin\theta)_{xxx}(\cos\theta)_{x}^2 \\
		 &+8(\sin\theta)_{xx}(\cos\theta)_{x}(\cos\theta)_{xx}+8(\sin\theta)_{xx}\cos\theta(\cos\theta)_{xxx}+8(\sin\theta)_{xxx}\cos\theta(\cos\theta)_{xx}+8(\sin\theta)_{xxx}(\cos\theta)_{x}^2\\
		 &+8(\sin\theta)_{xxx}(\cos\theta)_{x}^2+8(\sin\theta)_{xxx}\cos\theta(\cos\theta)_{xx}+8(\sin\theta)_{xxxx}\cos\theta(\cos\theta)_{x}\bigg] -\frac{1}{4}\alpha_{1}(t)(\sin\theta)_{xxx}\\
		 &-\frac{\alpha_3(t)}{2}\frac{1}{\sin\theta}\bigg[\frac{\bigg(\int^s\sin\theta(s_1,t)\mathrm{d}s_1\bigg)^2}
{\cos\theta}\bigg]_{x}=0.\label{72}
	\end{split}
\end{eqnarray}

Introduce a new dependent variable $v(s,t)=\int^s \sin \theta(s_1,t){\rm{d}} s_1$ (note $\tan \theta =v_x$, $\sin \theta= \frac{v_x}{\sqrt{1+v_x ^2}},
\cos \theta= \frac{1}{\sqrt{1+v_x ^2}}$). Then Eq.\eqref{72} is transformed to
\begin{equation}
	\begin{split}
		 &v_{xt'}-\frac{\alpha_{0}\left(t\right)}{16\left(1+v_x^2\right)^\frac{13}{2}}\Bigg[v_{xxxxxx}\left(1+v_x^2\right)^{4}-20v_{x}v_{xx}v_{xxxxx}\left(1+v_x^2\right)^{3}-35v_{xxxx}\Bigg(v_{xxx}\left(v_x^3+v_x\right)-\frac{39}{7}v_{xx}^2 \\
		 &(v_x^2-\frac{1}{6})\Bigg)\left(1+v_x^2\right)^{2}+270v_{xx}\Bigg(\left(1+v_x^2\right)^2(v_x^2-\frac{1}{6})v_{xxx}^2-\frac{77}{18}v_{x}v_{xxx}v_{xx}^2(v_x^2-\frac{1}{2})\left(1+v_x^2\right) \\
		 &+\frac{28}{9}v_{xx}^4\left(v_x^4-v_x^2+\frac{1}{16}\right)\Bigg)\Bigg]-\frac{1}{4}\alpha_{1}\left(t\right)\left(\frac{v_{x}}{\sqrt{1+v_x^2}}\right)_{xxx}-\alpha_3(t)\left(v+\frac{1}{6}\left(v^3\right)_{xx}\right)=0,
	\end{split}
\end{equation}
which is just the WKI-SP$^{(2,1)}$ equation \eqref{WKI-SP21}.

\subsection{Hodograph transformation between MKdV-SG$^{(1,2)}$ and WKI-SP$^{(1,2)}$ equation}
For $n=1,m=2$, the MKdV-SG$^{(1,2)}$ equation is given by

\begin{eqnarray}
q_{t}-\frac{1}{4}\bigg[\bigg(\frac{a_{1,s}}{q}\bigg)_{s}+4a_{1}q\bigg]_{s}+\frac{a_{2s}}{q}
=0.
\end{eqnarray}

By solving the recursive formulas \eqref{43} and \eqref{44} with $q=-\frac{\theta_{s}}{2}$, we get

\begin{eqnarray}
a_0=\alpha_0(t),\quad a_1=\frac{1}{2}\alpha_0(t)q^2+\alpha_1(t),\quad a_3=\frac{1}{4}\alpha_3(t)\cos{\theta },\quad
\end{eqnarray}
and
\begin{equation}
	\begin{split}
		&\frac{a_{2s}}{q}=C_{1}\sin{\theta}+C_{2}\cos{\theta}+2\sin{\theta}\int^{s} \sin \theta(s,t)\left(\int^{s_1}\sin \theta(s_2,t)\mathrm{d}s_2\right) \mathrm{d}s_1 \\
		&+2\cos{\theta}\int^{s}\cos \theta(s_1,t)\left(\int^{s_1}\sin \theta(s_2,t)\mathrm{d}s_2 \right) \mathrm{d}s_1,
	\end{split}
\end{equation}
with $q=-\frac{\theta_{s}}{2}$.

Then the MKdV-SG$^{(1,2)}$ equation \eqref{58} is rewritten as
\begin{equation}\label{77}
\begin{split}
&\theta _t=\frac{1}{4}\alpha_0(t)\theta _{sss}+\frac{1}{8}\alpha_0(t)\theta _{s}^3+\alpha_1(t)\theta_s+2C_1\int^s\sin{\theta}(s_1,t)\mathrm{d}s_1\\
&+2C_2\int^s\cos{\theta }(s_1,t)\mathrm{d}s_1+\frac{2}{3}\biggl(\int^s\sin{\theta}(s_1,t)\mathrm{d}s_1\biggr)^3\\
&+4\int^s\cos{\theta}(s_1,t)\biggl[\int^{s_1}\sin{\theta}(s_2,t)\biggl(\int^{s_2}\sin{\theta }(s_3,t)\mathrm{d}s_3\biggr)\mathrm{d}s_2\biggr]\mathrm{d}s_1.
\end{split}
\end{equation}


The corresponding conservation law of Eq.\eqref{77} is
\begin{equation}
	\begin{split}
		 &(\cos\theta)_t=\Bigg[-\frac{1}{8}\alpha_0(t)\left(2\theta_{ss}-\theta_s^2\cos{\theta}\right)+\alpha_1(t)\cos{\theta}-C_{1}\left(\int^{s}\sin{\theta}(s_1,t)\mathrm{d}s_1\right)^2 \\
		 &-\frac{1}{6}\left(\int^{s}\sin{\theta}(s_1,t)\mathrm{d}s_1\right)-\int^{s}A\mathrm{d}s_{1} \Bigg]_s
	\end{split}\label{78}
\end{equation}
with
\begin{equation}
	\begin{split}
		 &A=2C_2\sin{\theta}\int^{s}\cos{\theta}(s_1,t)\mathrm{d}s_1+4\sin{\theta}\int^{s}\cos{\theta}(s_1,t)\left[\int^{s_1}\cos{\theta}(s_2,t)\left(\int^{s_2}\cos{\theta}(s_3,t)\mathrm{d}s_3\right)\mathrm{d}s_2\right]\mathrm{d}s_1.
	\end{split}
\end{equation}

The same hodograph transformation \eqref{3-29} leads \eqref{78} to
\begin{eqnarray}
	\frac{\partial}{\partial s}=\cos \theta\frac{\partial}{\partial x}\label{79},
\end{eqnarray}
\begin{eqnarray}
	\begin{split}
		&\frac{\partial}{\partial t}=\Biggl[-\frac{1}{8}\alpha_0(t)(2\theta _{ss}\sin{\theta }-\theta _{s}^2\cos{\theta })+\alpha _1(t)\cos{\theta }-C_1\biggl(\int^s \sin{\theta }(s_1,t)\mathrm{d}s_1\biggr)-\frac{1}{6}\biggl(\int^s\sin{\theta }(s_1,t)\mathrm{d}s_1\biggr)^4\\
		&-\int^sA\mathrm{d}s\Biggr]\frac{\partial}{\partial x}+\frac{\partial}{\partial t'}.
	\end{split}\label{80}
\end{eqnarray}

Applying \eqref{79} and \eqref{80} to Eq.\eqref{78}, we obtain
\begin{eqnarray}
	(\tan(\theta ))_{t'}-\frac{1}{4}(\sin{\theta })_{xxx}-\Biggl[\frac{C_{1}\left(\int^{s}\sin{\theta}(s_1,t)\mathrm{d}s_1\right)^2+\frac{1}{6}\left(\int^{s}\sin{\theta}(s_1,t)\mathrm{d}s_1\right)^4-\int^{s}A\mathrm{d}s_{1}}{\cos{\theta }}\Biggr]_x\frac{1}{\sin{\theta }}=0.\label{81}
\end{eqnarray}

Introduce a new dependent variable $v(s,t)=\int^s \sin \theta(s_1,t){\rm{d}} s_1$ (note $\tan \theta =v_x$, $\sin \theta= \frac{v_x}{\sqrt{1+v_x ^2}},
\cos \theta= \frac{1}{\sqrt{1+v_x ^2}}$).
Then Eq.\eqref{81} is transformed to
\begin{equation}
	\begin{split}
		 &v_{xt'}=\frac{1}{4}\alpha_0(t)\left(\frac{v_x}{\sqrt{1+v_x^2}}\right)_{xxx}+C_{1}v^{2}v_{xx}+2C_{2}xv_{xx}+2C_{2}v_{xx}\int^{x}v\mathrm{d}x' +4v_{xx}\int^{x}v'_x\left[\left(\int^{x'}v\mathrm{d}x'''\right)\mathrm{d}x''\right]\mathrm{d}x'\\
		 &+2C_{1}v(1+v_x^2)+\frac{2}{3}v^3(1+v_x^2)+\frac{1}{6}v^{4}v_{xx}+2C_{2}x(1+v_x^2)
+4(1+v_x^2)\int^{x}\left(\int^{x'}v\mathrm{d}x''\right)\mathrm{d}x'.
	\end{split}
\end{equation}
We set $v=u_{xx}$ with $C_1=C_2=0$ and obtain
\begin{equation}
	\begin{split}
		 &u_{xxxt'}=\frac{1}{4}\alpha_0(t)\left(\frac{u_{xxx}}{\sqrt{1+u_{xxx}^2}}\right)_{xxx}+
\frac{1}{30}(u_{xx}^{5})_{xx}+\left[(4uu_{xx}-2u_x^2)u_{xxx}\right]_x+\frac{2}{3}u_{xx}^3+4u.
	\end{split}
\label{86}
\end{equation}
Now Eq.\eqref{86} is just the  WKI-SP$^{(1,2)}$ equation.
\\

\textbf{Remark 1:} We believe that the hodograph transformation defined by \eqref{3-29} and \eqref{3-33} is not only true of the three pairs of equations given above, but also valid for the the whole MKdV-SG and the WKI-SP hierarchy.

\section{Loop soliton solutions of the WKI-SP$^{(1,1)}$ equation}

For simplicity, we rewrite the MKdV-SG$^{(1,1)}$ equation  and the WKI-SP$^{(1,1)}$ equation as
\begin{eqnarray}
\theta_{ts}+\alpha\bigg[\frac{3}{2}\theta_s^2\theta_{ss}+\theta_{ssss}\bigg]=\beta \sin(\theta)
\label{MKdV-SG}
\end{eqnarray}
 and
\begin{eqnarray}
u_{xt}+\alpha \bigg(\frac{u_x}{\sqrt{1+u_x ^2}}\bigg)_{xxx}=\beta\bigg(u+\frac{1}{6}(u^3)_{xx}\bigg)
\label{WKI-SP}
\end{eqnarray}
with $\alpha$ and $\beta$ being arbitrary real constants.

In Ref.\cite{ZhangDJ2002,ZhangDJ2004}, the authors presented the following  $N$-soliton solution for
the MKdV-SG$^{(1,1)}$ equation \eqref{MKdV-SG}:
\begin{eqnarray}
\theta=2i\ln \frac{{f}^{*}}{f}
\label{4-39}
\end{eqnarray}
with
\begin{eqnarray}
&&f=f_N=\sum _{\mu=0,1} \exp \bigg[\sum_{j=1}^N \mu_j(\xi_j+\frac{\pi}{2}i)+\sum_{1\leq j<l\leq N} ^N \mu_j\mu_l A_{jl}\bigg],\label{4-40}\\
&& \xi_j=k_j s+\omega_j t+\xi_{j0},\quad \quad \omega_j=\frac{\beta}{k_j}-\alpha k_j^3, \quad \quad {\rm {e}}^{A_{jl}}=\frac{(k_j-k_l)^2}{(k_j+k_l)^2},\quad \quad k_i\neq \pm k_j.
\label{4-41}
\end{eqnarray}
where the summation of $\mu \equiv \{\mu_1, \mu_2, \ldots, \mu_N\}$
should be done for all permutations of $\mu_j=0,1, j=1,2, \ldots, N$.
Here${f}^{*}$ is the conjugate of $f$.

Due to the hodograph transformation \eqref{3-29} and \eqref{3-33}, the loop soliton
solutions of the WKI-SP$^{(1,1)}$ equation are represented by the parametric form as

\begin{eqnarray}
&&u=\frac{1}{\beta}[\theta_{t}+\alpha(\frac{1}{2}\theta_s^3+\theta_{sss})],\quad(\theta=2i\ln \frac{\bar{f}}{f}),\label{4-42-u}\\
&&x=s-\frac{2}{\beta}\bigg[(\ln{f\bar{f}})_t+\alpha(\ln{f\bar{f}})_{sss}+6\alpha\int ^s \bigg((\ln f)^2_{ss}+(\ln \bar{f})^2_{ss}\bigg)\mathrm{d}s\bigg]+x_0,
\label{4-42-x}
\end{eqnarray}
where $f$ satisfies \eqref{4-40}.

\subsection{One-loop soliton solution}

In the case of $N=1$, we have
\begin{eqnarray}
f=1+{\rm {ie}}^{\xi_1},\quad  \xi_1=k_1 s+\bigg(\frac{\beta}{k_1}-\alpha k_1^3\bigg) t+\xi_{10}.
\label{4-43}
\end{eqnarray}
Substituting \eqref{4-43} into \eqref{4-42-u} and \eqref{4-42-x}  leads to
\begin{eqnarray}
	&&u(s,t)=\frac{2}{k_1}  {\rm {sech}} \xi_1,\label{93}\\
	&&x(s,t)=s-\frac{2}{k_1}\tanh\xi_1+d+\frac{2}{k_1}, \label{94}
\end{eqnarray}
which is the well-known one soliton solution and  depicted in Figures 1 and 2.
The positive and negative $k_1$  correspond to loop
soliton and antiloop soliton, respectively. Most of the time, we call both of them
loop soliton for convenience.

\begin{figure}[H]
	\centering
	
		\subfigure[]{
		\begin{minipage}[t]{0.3\linewidth}
		\centering
		\includegraphics*[width=1.5in]{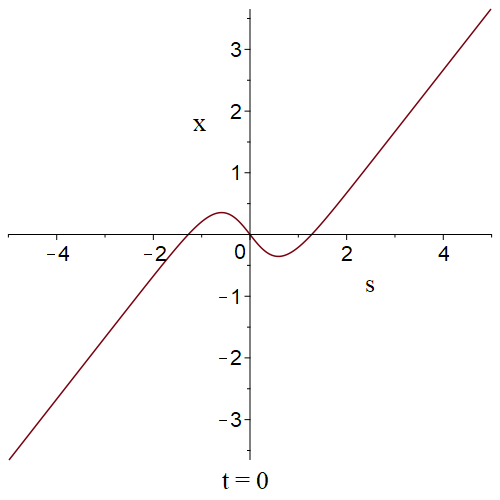}%
		 \end{minipage}%
		}%
		\subfigure[]{
		\begin{minipage}[t]{0.3\linewidth}
		\centering
		\includegraphics*[width=1.5in]{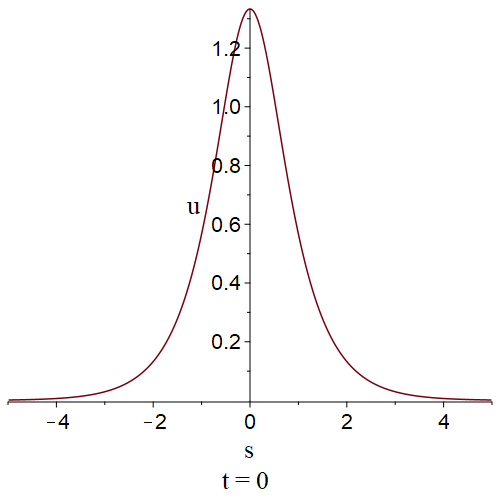}%
		\end{minipage}%
		}%
		\subfigure[]{
		 \begin{minipage}[t]{0.3\linewidth}
		 \centering
		 \includegraphics*[width=1.5in]{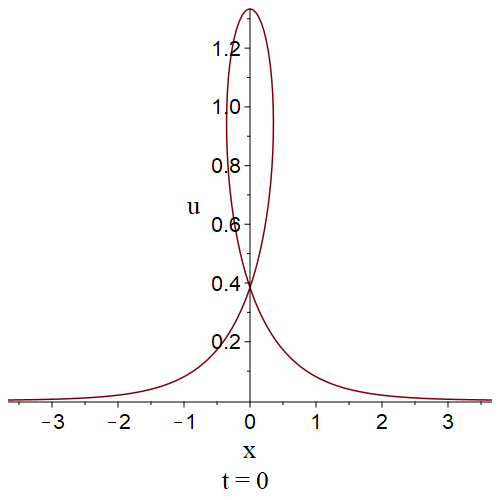}%
		 \end{minipage}
		}%
		\caption{a, b, and c are the profiles of x as a function of s, u as a function of s, and one-loop soliton solution of the WKI-SP$^{(1,1)}$ equation with parameters $\alpha=2$, $\beta=2$, $k_1=1.5$, $\xi_{10}=0$ and $d_{1}=-\frac{4}{3}$ respectively.}
\end{figure}

\begin{figure}[H]
	\centering
	
		\subfigure[]{
		\begin{minipage}[t]{0.3\linewidth}
		\centering
		\includegraphics*[width=1.5in]{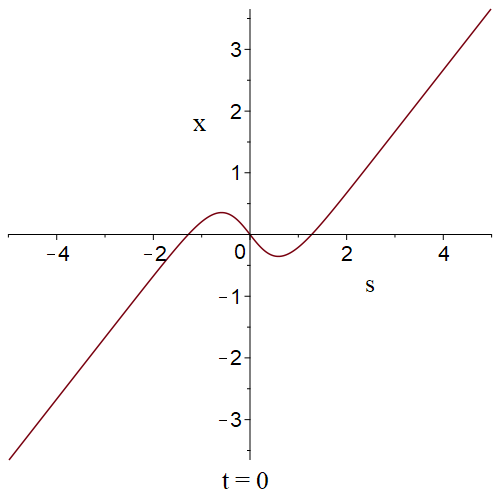}%
		 \end{minipage}%
		}%
		\subfigure[]{
		\begin{minipage}[t]{0.3\linewidth}
		\centering
		\includegraphics*[width=1.5in]{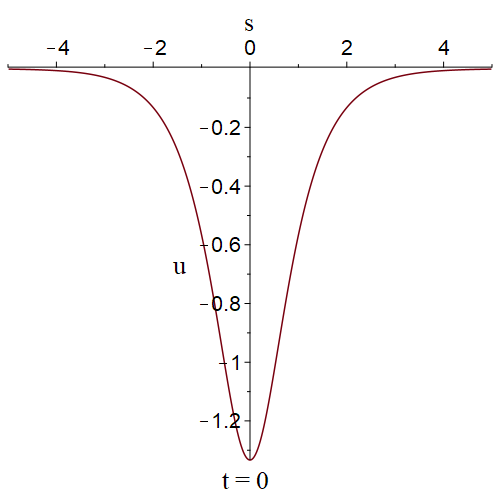}%
		\end{minipage}%
		}%
		\subfigure[]{
		 \begin{minipage}[t]{0.3\linewidth}
		 \centering
		 \includegraphics*[width=1.5in]{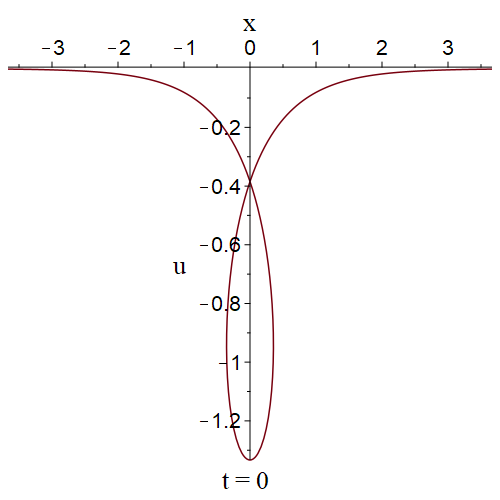}%
		 \end{minipage}
		}%
		\caption{a, b, and c are the profiles of x, as a function of s, u as a function of s, and one-antiloop soliton solution of the WKI-SP equation with parameters $\alpha=2$, $\beta=2$, $k_1=-1.5$, $\xi_{10}=0$ and $d_{1}=\frac{4}{3}$ respectively.}
\end{figure}

To clarify the motion of the loop in the $(x,t)$ coordinate system, we rewrite \eqref{94} into the form
\begin{eqnarray}
	x-c_1t+x_{10}=\frac{\xi_1}{k_1}-\frac{2}{k_1}\tanh\xi_1+d+\frac{2}{k_1},\label{4-45}
\end{eqnarray}
with $c_1={\alpha}k_{1}^{2}-\frac{\beta}{k_{1}^2}$ and $x_{10}=\frac{\xi_{10}}{k_1}$.

Here,  $c_1$ denotes the  velocity of the loop soliton.
For the given nonzero $\alpha$ and $\beta$, it shows that the loop soliton could propagate to either the left (i.e.,negative x direction) or the right (i.e., positive x direction). This is different from the SP equation and the WKI equation. The soliton of the SP equation and the WKI equation can only move in one direction of $x$-axis.  From \eqref{93}, we see the amplitude (defined by $A_1$) of the loop soliton is $\frac{2}{k_1}$. Then we have
\begin{eqnarray}
c_1=\frac{4\alpha}{A_{1}^2}-\frac{{\beta}A_1^2}{4}.\label{96}
\end{eqnarray}
Interestingly, the above expression shows the loop solitons of the WKI-SP$^{(1,1)}$ equation have abundant solitonic behaviors. On the one hand, the large loop can
move more rapidly than the small  loop. On the other hand, the small loop can also move more rapidly than the large  loop, which is different from  the typical solitonic movement.

\subsection{Interaction of two loop solitons}

For the two-loop soliton solution, \eqref{4-40} gives
\begin{eqnarray}
	f=1+{\rm ie}^{\xi_1}+{\rm ie}^{\xi_2}-{\gamma}{\rm e}^{\xi_1+\xi_2}\label{4-46}
\end{eqnarray}
with
\begin{eqnarray}
	\gamma=\bigg(\frac{k_1-k_2}{k_1+k_2}\bigg)^2.
	\label{4-47}
\end{eqnarray}

Due to \eqref{4-42-u} and \eqref{4-42-x}, the parametric representation of the solution reads
\begin{eqnarray}
	 &&u(s,t)=\frac{2\sqrt{\gamma}}{k_{1}k_2}\frac{(k_1+k_2)\cosh\Psi_{1}\cosh\Psi_2+(k_1-k_2)\sinh\Psi_{1}\sinh\Psi_2}{\cosh^{2}\Psi_1+{\gamma}\sinh^{2}\Psi_2},\\
	 &&x(s,t)=s+\frac{1}{k_{1}k_2}\frac{(k_1-k_2)\sinh2\Psi_1-\gamma(k_1+k_2)\sinh2\Psi_2}{\cosh^{2}\Psi_1+\gamma\sinh^{2}\Psi_2}+\frac{2(k_1+k_2)}{k_1k_2}+d,\label{4-48}
\end{eqnarray}
where we have put
\begin{eqnarray}
\Psi_1=\frac{1}{2}(\xi_1-\xi_2),\quad
\Psi_2=\frac{1}{2}(\xi_1+\xi_2)+\frac{1}{2}\ln \gamma \label{4-49}
\end{eqnarray}
for simplicity.

The two-loop soliton has abundant solitonic behaviors. Two loops can move in
the same direction of the $x$-axes while they could move towards each other in the opposite directions. It could be the fast small loop chasing the slow high one, and it might be the fast high loop chasing the slow small one. To figure out what each parameter does, we rewrite the
amplitude and the velocity as follows
\begin{eqnarray}
A_i=\frac{2}{k_i},	\quad  c_i={\alpha}k_i^2-\frac{\beta}{k_i^2},  \quad (i=1,2).
\end{eqnarray}
A table is listed to intuitively show the chasing or the meeting movement.
Since $|k_1|\neq|k_2|$, we assume $|k_1|>|k_2|$ without generality. In the following table,
$v_{low}$ represents the speed of the small loop soliton while  $v_{high}$ represents the speed of the large  loop soliton.
\begin{table}[H]
	\caption{The role of each parameter plays in the movement of two-loop soliton}
	\begin{center}
		 \begin{tabular}{|p{0.5cm}<{\centering}|p{3cm}<{\centering}|p{2cm}<{\centering}|p{2cm}<{\centering}|p{2cm}<{\centering}|p{2cm}<{\centering}|p{1.5cm}<{\centering}|}\hline
			\multicolumn{2}{|c|}{\multirow{2}{*}{$ $}} & \multicolumn{4}{c|}{one loop chases another loop} & \multirow{2}{*}{two loops}\\
			\cline{3-6}
			\multicolumn{2}{|c|}{~} & both move to the positive $x$-axis & both move to the negative $x$-axis & $v_{low}>v_{high}$ & $v_{low}<v_{high}$ &  meet each other\\
			\cline{1-7}
			\multirow{3}{*}{~}  & $\sqrt[4]{\frac{\beta}{\alpha}}<|k_2|<|k_1|$ & $\checkmark$ $(\alpha>0,\beta>0)$ & $\checkmark$ $(\alpha<0,\beta<0)$ & $\checkmark$ &  & \\
			\cline{2-7}
			\rotatebox{90}{$\alpha\beta>0$} & $|k_2|<|k_1|<\sqrt[4]{\frac{\beta}{\alpha}}$  & $\checkmark$ $(\alpha<0,\beta<0)$ & $\checkmark$ $(\alpha>0,\beta>0)$ &  & $\checkmark$ & \\
			\cline{2-7}
			& $|k_2|<\sqrt[4]{\frac{\beta}{\alpha}}<|k_1|$ &  &  &  &  & $\checkmark$\\
			\cline{1-7}
		
			\multirow{2}{*}{~} & $\frac{\sqrt{-\alpha\beta}}{|\alpha||k_1|}<|k_2|<|k_1|$ & $\checkmark$ $(\alpha>0,\beta<0)$ & $\checkmark$ $(\alpha<0,\beta>0)$ & $\checkmark$ &  & \\
			\cline{2-7}
			\rotatebox{90}{$\alpha\beta<0$}& $|k_2|<min\{|k_1|,\frac{\sqrt{-\alpha\beta}}{|\alpha||k_1|}\}$ & $\checkmark$ $(\alpha>0,\beta<0)$ & $\checkmark$ $(\alpha<0,\beta>0)$ &  & $\checkmark$ & \\
			\cline{1-7}
		\end{tabular}
	\end{center}
\end{table}
Some diverse figures are worked out to exhibit different
solitonic movements.
In Figures 3 and 4, both the two loop solitons are moving to the right. There are both
a small but fast loop soliton chasing a larger and slower loop soliton.
Two loop solitons with dissimilar amplitudes are shown in Figures 3, in which
a small and fast loop soliton maybe observed traveling around a
larger and slower loop soliton.
Two loop solitons with similar amplitudes ($k_1\approx k_2$) are shown in Figures 4, in which
the loops do not over lap and they just seems to exchange their amplitudes during the period of the nonlinear interaction.

\begin{figure}[H]
	\centering
	
		\subfigure[]{
		\begin{minipage}[t]{0.25\linewidth}
		\centering
		\includegraphics*[width=1.5in]{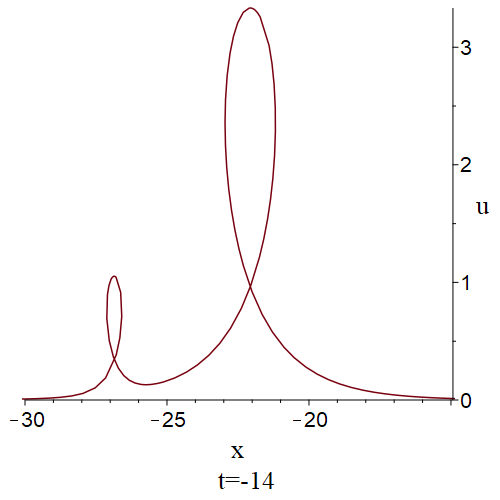}%
		 \end{minipage}%
		}%
		\subfigure[]{
		\begin{minipage}[t]{0.25\linewidth}
		\centering
		\includegraphics*[width=1.5in]{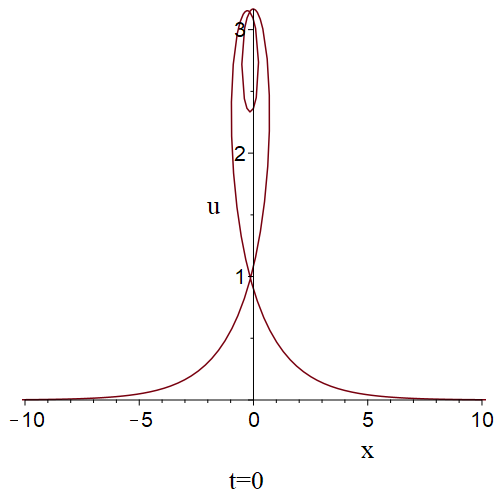}%
		\end{minipage}%
		}%
		\subfigure[]{
		 \begin{minipage}[t]{0.25\linewidth}
		 \centering
		 \includegraphics*[width=1.5in]{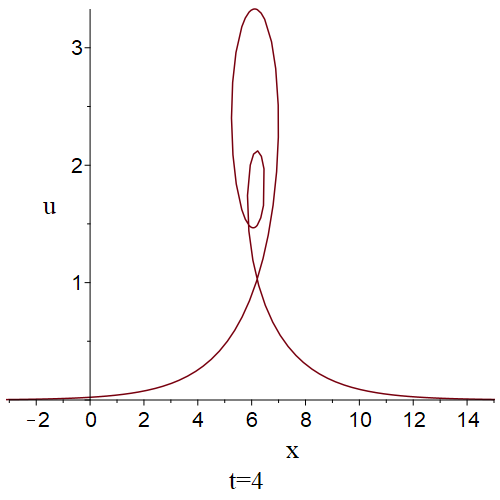}%
		 \end{minipage}
		}%
		\subfigure[]{
		\begin{minipage}[t]{0.25\linewidth}
		\centering
		 \includegraphics*[width=1.5in]{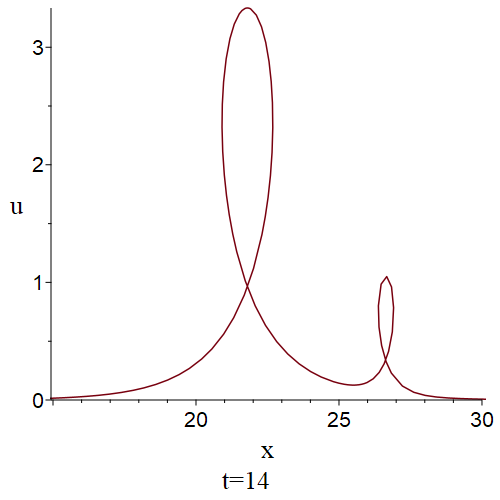}%
		\end{minipage}
		}%
		\centering
		\caption{The interaction process for the pursuing of two loop solitons with $\alpha=0.5$, $\beta=-0.5$, $k_1=2$, $k_2=0.6$, $\xi_{10}=-1.5$, $\xi_{20}=0$ and $x_{0}=-5.5$.}
\end{figure}

\begin{figure}[H]
	\centering
		\subfigure[]{
		\begin{minipage}[t]{0.25\linewidth}
		\centering
		\includegraphics*[width=1.5in]{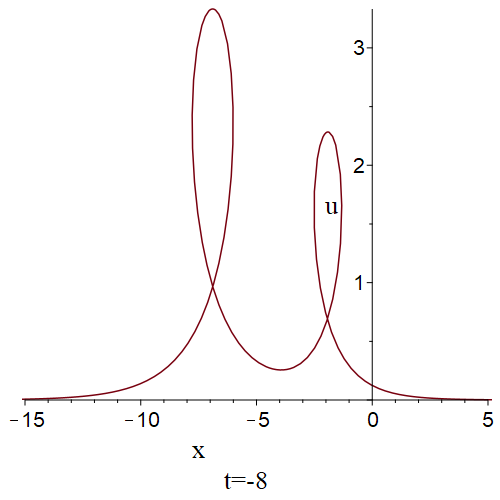}%
		 \end{minipage}%
		}%
		\subfigure[]{
		\begin{minipage}[t]{0.25\linewidth}
		\centering
		\includegraphics*[width=1.5in]{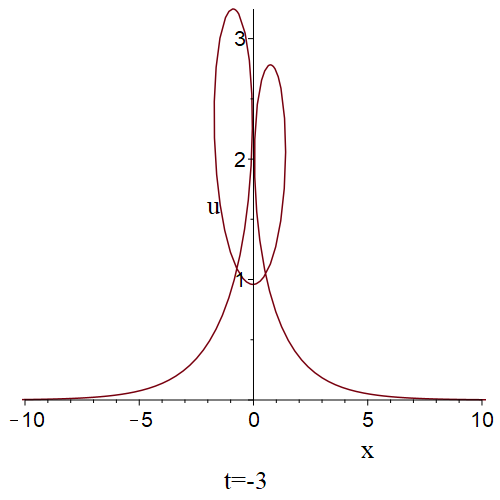}%
		\end{minipage}%
		}%
		\subfigure[]{
		 \begin{minipage}[t]{0.25\linewidth}
		 \centering
		 \includegraphics*[width=1.5in]{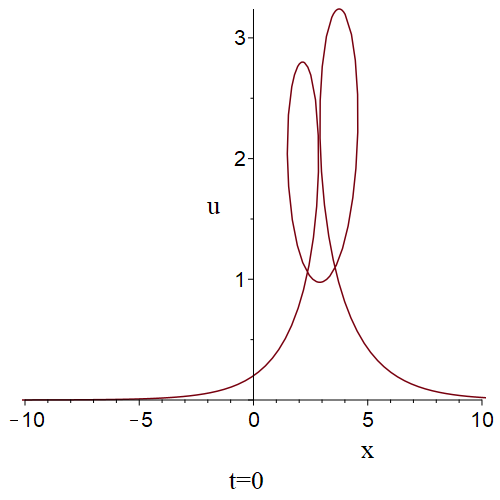}%
		 \end{minipage}
		}%
		\subfigure[]{
		\begin{minipage}[t]{0.2\linewidth}
		\centering
		 \includegraphics*[width=1.5in]{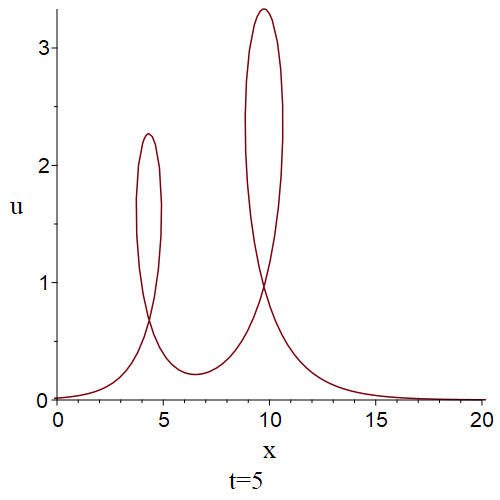}%
		\end{minipage}
		}%
		\centering
		\caption{The interaction process for the pursuing of two loop solitons with $\alpha=-0.5$, $\beta=-0.5$, $k_1=0.9$, $k_2=0.6$, $\xi_{10}=1$, $\xi_{20}=0$ and $x_{0}=-5$.}
\end{figure}

Figures 5 exhibit two loop solitons are moving to the left. Here a small and slower loop
soliton catches up with a larger and fast loop soliton.

\begin{figure}[H]
	\centering
		\subfigure[]{
		\begin{minipage}[t]{0.25\linewidth}
		\centering
		\includegraphics*[width=1.5in]{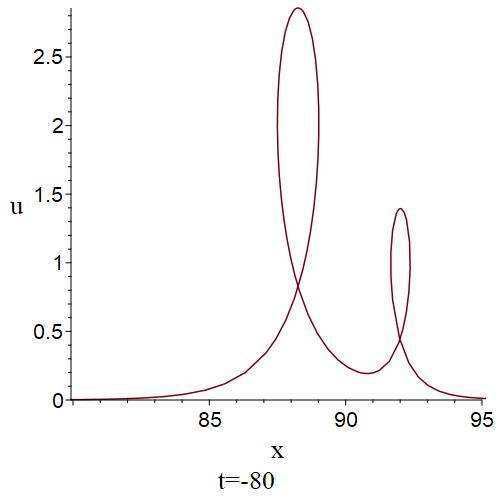}%
		 \end{minipage}%
		}%
		\subfigure[]{
		\begin{minipage}[t]{0.25\linewidth}
		\centering
		\includegraphics*[width=1.5in]{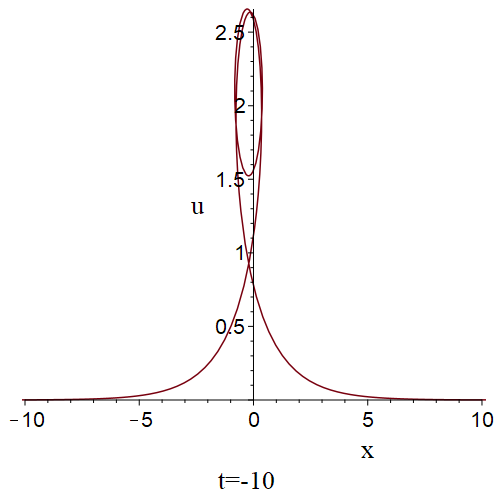}%
		\end{minipage}%
		}%
		\subfigure[]{
		 \begin{minipage}[t]{0.25\linewidth}
		 \centering
		 \includegraphics*[width=1.5in]{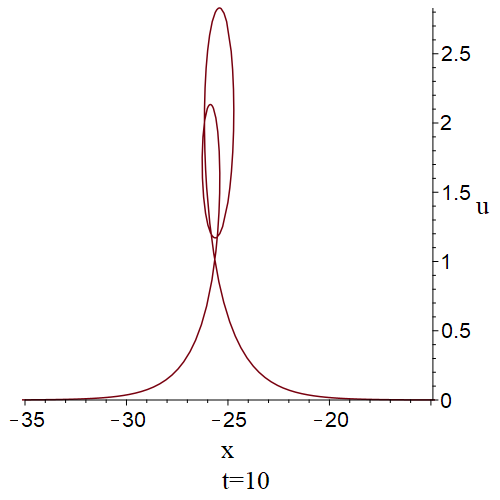}%
		 \end{minipage}
		}%
		\subfigure[]{
		\begin{minipage}[t]{0.2\linewidth}
		\centering
		 \includegraphics*[width=1.5in]{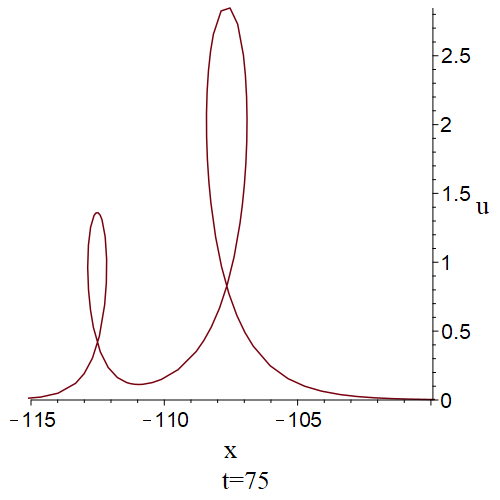}%
		\end{minipage}
		}%
		\centering
		\caption{The interaction process for the collision of two loop solitons  with $\alpha=-0.5$, $\beta=0.5$, $k_1=1.5$, $k_2=0.7$, $\xi_{10}=0$, $\xi_{20}=0$ and $x_{0}=-18.5$.}
\end{figure}

Figures 6 illustrate one loop soliton and one antiloop soliton  move towards each other in  the
opposite directions. The antiloop travels to the right and  the loop
travels to the left. As time goes, two solitons merge and
then they separate each other with leaving the original wave
profiles.

\begin{figure}[H]
	\centering
	
		\subfigure[]{
		\begin{minipage}[t]{0.3\linewidth}
		\centering
		\includegraphics*[width=1.5in]{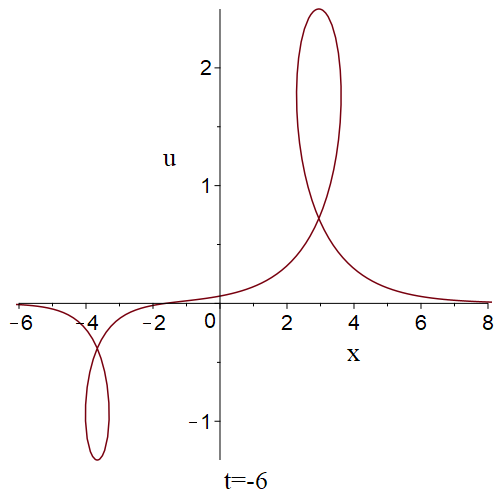}%
		 \end{minipage}%
		}%
		\subfigure[]{
		\begin{minipage}[t]{0.3\linewidth}
		\centering
		\includegraphics*[width=1.5in]{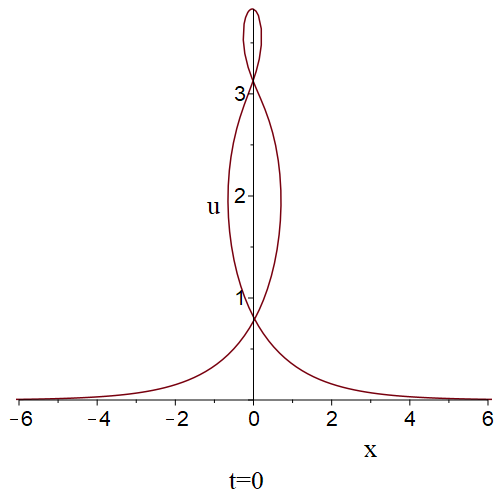}%
		\end{minipage}%
		}%
		\subfigure[]{
		 \begin{minipage}[t]{0.3\linewidth}
		 \centering
		 \includegraphics*[width=1.5in]{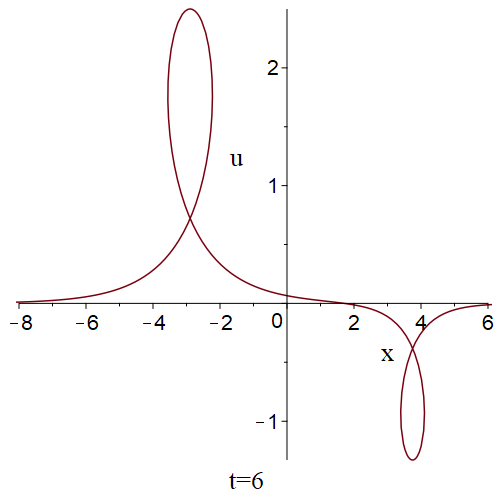}%
		 \end{minipage}
		}%
		\caption{The interaction process for the collision of one loop and one antiloop soliton with $\alpha=0.5$, $\beta=0.5$, $k_1=-1.5$, $k_2=0.8$, $\xi_{10}=-3.4$, $\xi_{20}=0$ and $x_{0}=0.3$.}
\end{figure}

\subsection{Interaction of three loop solitons}

For the three-loop soliton case, formulas \eqref{4-40} reduce to
\begin{eqnarray}
	f=1+{\rm ie}^{\xi_1}+{\rm ie}^{\xi_2}+{\rm ie}^{\xi_3}-{\gamma_{12}}{\rm e}^{\xi_1+\xi_2}
-{\gamma_{13}}{\rm e}^{\xi_1+\xi_3}-{\gamma_{23}}{\rm e}^{\xi_2+\xi_3}-{\rm i} {\gamma_{123}}{\rm e}^{\xi_1+\xi_2+\xi_3}
\label{4-132}
\end{eqnarray}
with
\begin{eqnarray}
	\gamma_{ij}=\bigg(\frac{k_i-k_j}{k_i+k_j}\bigg)^2, \quad \gamma_{123}=\gamma_{12}\gamma_{13}\gamma_{23}.
	\label{4-133}
\end{eqnarray}

By virtue of  \eqref{4-42-u} and \eqref{4-42-x}, the three-loop soliton in the parametric form is provided as follows
\begin{eqnarray}
	 &&u(s,t)=\frac{2\Xi_1}{\Xi},\\
	 &&x(s,t)=s+\frac{2\Xi_2}{\Xi}+\frac{2(k_1k_2+k_1k_3+k_2k_3)}{k_1k_2k_3}+d.\label{4-134}
\end{eqnarray}
where
\begin{eqnarray}
	\begin{split}
		 \Xi&=k_{1}k_{2}k_{3}[2(\gamma_{12}\sinh^{2}\Psi_3+\gamma_{13}\sinh^{2}\Psi_2+\gamma_{23}\sinh^{2}\Psi_1)+2(1-\gamma_{13})\sqrt{\gamma_{12}\gamma_{23}}\cosh(\Psi_1+\Psi_3)\\
		 &+2(1-\gamma_{23})\sqrt{\gamma_{12}\gamma_{13}}\cosh(\Psi_2+\Psi_3)+2(1-\gamma_{12})\sqrt{\gamma_{13}\gamma_{23}}\cosh(\Psi_1+\Psi_2)\\
		&+\gamma_{123}\cosh(2(\Psi_{1}+\Psi_{2}+\Psi_{3}))+(\gamma_{12}+\gamma_{13}+\gamma_{23})]
	\end{split}
\end{eqnarray}

\begin{equation}
	\begin{split}
		\Xi_1=&\sum_{1 \leqslant m<n\leqslant 3}\gamma_{m,n}\bigg(\sum_{1 \leqslant i<j\leqslant 3}k_{i}k_{j}-2k_{m}k_{n}\bigg)-\gamma_{123}\sum_{1 \leqslant i<j\leqslant 3}k_{i}k_{j}\\
		 &+2k_{1}k_{2}\sqrt{\gamma_{13}\gamma_{23}}\bigg[\gamma_{12}\cosh(\Psi_{1}+\Psi_{2}+2\Psi_{3})+\cosh(\Psi_{1}-\Psi_{2})\bigg]\\
		 &+2k_{1}k_{3}\sqrt{\gamma_{12}\gamma_{23}}\bigg[\gamma_{13}\cosh(\Psi_{1}+2\Psi_{2}+\Psi_{3})+\cosh(\Psi_{1}-\Psi_{3})\bigg]\\
		 &+2k_{2}k_{3}\sqrt{\gamma_{12}\gamma_{13}}\bigg[\gamma_{23}\cosh(2\Psi_{1}+\Psi_{2}+\Psi_{3})+\cosh(\Psi_{2}-\Psi_{3})\bigg]
	\end{split}
\end{equation}

\begin{equation}
	\begin{split}
		 \Xi_2=&\gamma_{123}(k_1k_2+k_1k_3+k_2k_3)\sinh(2(\Psi_1+\Psi_2+\Psi_3))+\gamma_{12}(-k_1k_2+k_1k_3+k_2k_3)\sinh(2\Psi_3)\\
		 &+\gamma_{13}(k_1k_2-k_1k_3+k_2k_3)\sinh(2\Psi_2)+\gamma_{23}(k_1k_2+k_1k_3-k_2k_3)\sinh(2\Psi_1)\\
		 &+k_2k_3(\gamma_{23}-1)\sqrt{\gamma_{12}\gamma_{13}}\sinh(\Psi_2+\Psi_3)+k_1k_3(\gamma_{13}-1)\sqrt{\gamma_{12}\gamma_{23}}\sinh(\Psi_1+\Psi_3)\\
		&+k_1k_2(\gamma_{12}-1)\sqrt{\gamma_{13}\gamma_{23}}\sinh(\Psi_1+\Psi_2)
	\end{split}
\end{equation}

with
\begin{eqnarray}
	&&\Psi_1=\frac{1}{2}(-\xi_1+\xi_2+\xi_3)+\frac{1}{2}\ln(\gamma_{23}),\\
	&&\Psi_2=\frac{1}{2}(\xi_1-\xi_2+\xi_3)+\frac{1}{2}\ln(\gamma_{13}),\\
	&&\Psi_3=\frac{1}{2}(\xi_1+\xi_2-\xi_3)+\frac{1}{2}\ln(\gamma_{12}).
\end{eqnarray}




Two examples of such solution are plotted below (see Figures 7 and 8).
Figures 7 describe three loop solitons with different  speed  propagating to the positive direction. It shows rich catch-up phenomenon. Figures 8 describes two loop solitons colliding with one antiloop soliton.

\begin{figure}[H]
	\centering
		\subfigure[]{
		\begin{minipage}[t]{0.3\linewidth}
		\centering
		\includegraphics*[width=1.5in]{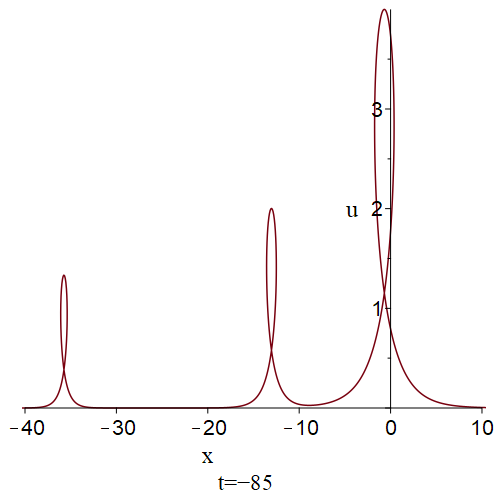}%
		 \end{minipage}%
		}%
		\subfigure[]{
		\begin{minipage}[t]{0.3\linewidth}
		\centering
		\includegraphics*[width=1.5in]{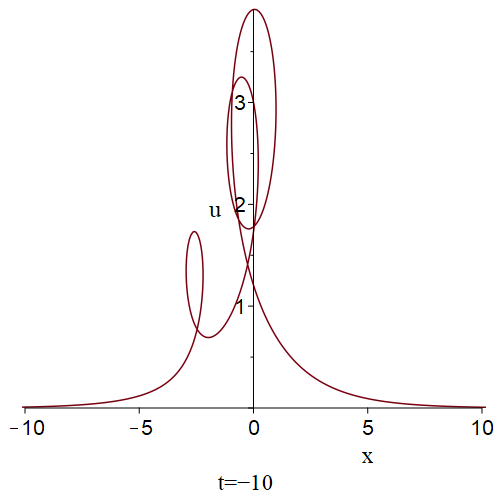}%
		\end{minipage}%
		}%
		\subfigure[]{
		 \begin{minipage}[t]{0.3\linewidth}
		 \centering
		 \includegraphics*[width=1.5in]{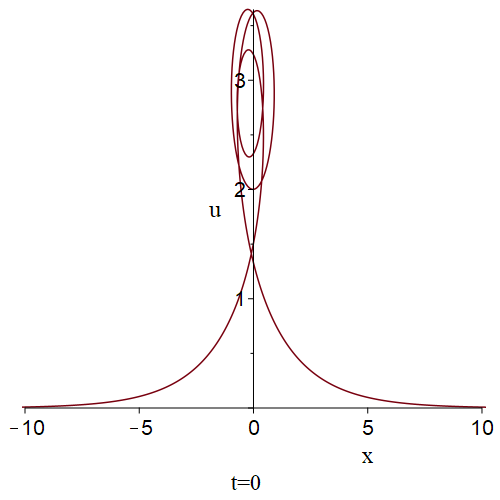}%
		 \end{minipage}
		}%

		\subfigure[]{
		\begin{minipage}[t]{0.3\linewidth}
		\centering
		 \includegraphics*[width=1.5in]{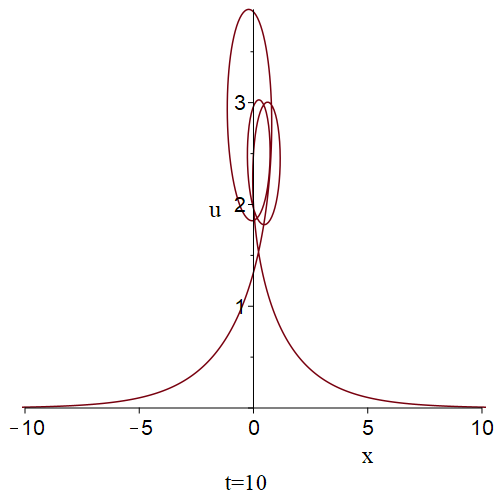}%
		\end{minipage}
		}%
		\centering
		\subfigure[]{
		\begin{minipage}[t]{0.3\linewidth}
		\centering
		 \includegraphics*[width=1.5in]{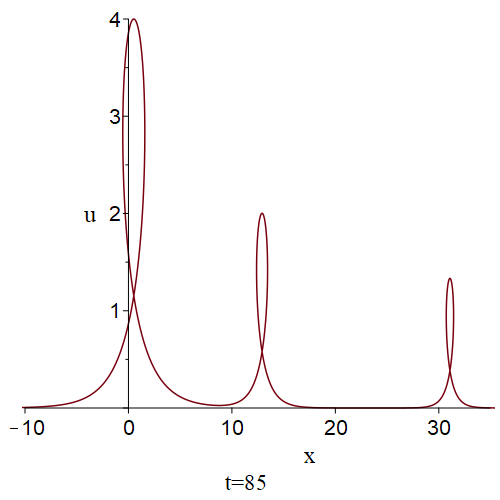}%
		\end{minipage}
		}%
		\centering
		\caption{The interaction process for three loop solitons with $\alpha=0.2$, $\beta=0.01$, $k_1=0.5$, $k_2=1$, $k_3=1.5$, $\xi_{10}=0$, $\xi_{20}=-0.9$, $\xi_{30}=0.3$ and $d=-11$.}
\end{figure}


\begin{figure}[H]
	\centering
		\subfigure[]{
		\begin{minipage}[t]{0.3\linewidth}
		\centering
		\includegraphics*[width=1.5in]{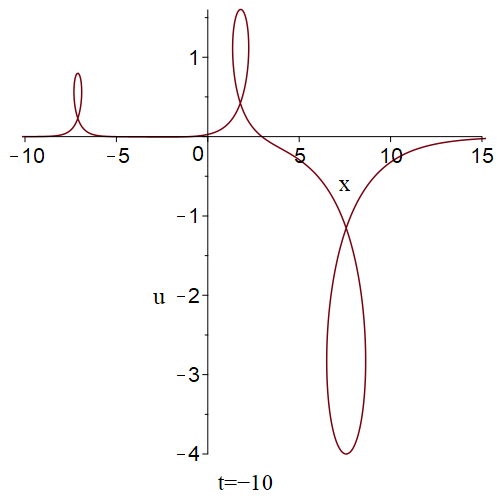}%
		 \end{minipage}%
		}%
		\subfigure[]{
		\begin{minipage}[t]{0.3\linewidth}
		\centering
		\includegraphics*[width=1.5in]{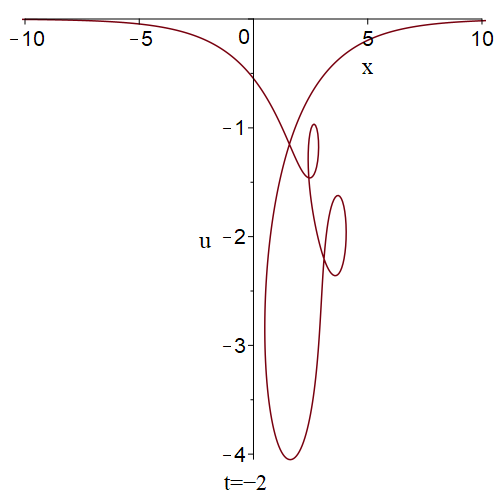}%
		\end{minipage}%
		}%
		\subfigure[]{
		 \begin{minipage}[t]{0.3\linewidth}
		 \centering
		 \includegraphics*[width=1.5in]{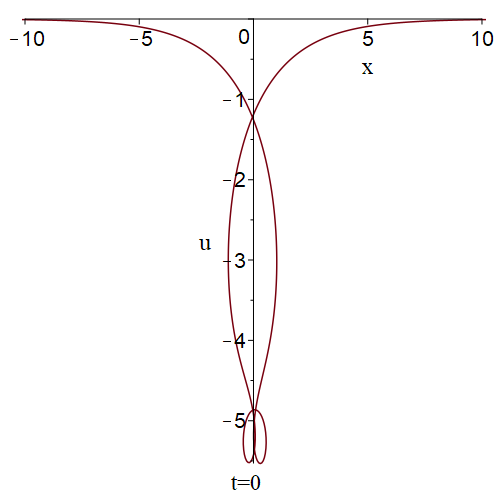}%
		 \end{minipage}
		}%

		\subfigure[]{
		\begin{minipage}[t]{0.3\linewidth}
		\centering
		 \includegraphics*[width=1.5in]{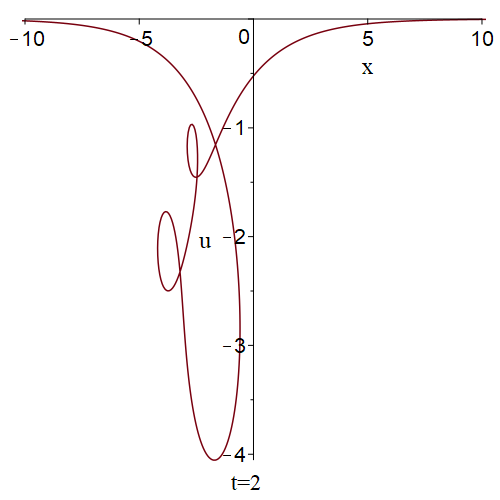}%
		\end{minipage}
		}%
		\centering
		\subfigure[]{
		\begin{minipage}[t]{0.3\linewidth}
		\centering
		 \includegraphics*[width=1.5in]{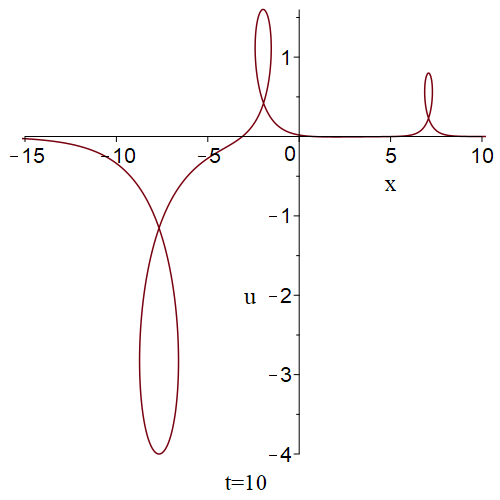}%
		\end{minipage}
		}%
		\centering
		\caption{The interaction process for two loops and one antiloop soliton with $\alpha=0.2$, $\beta=0.2$, $k_1=1.2$, $k_2=2.5$, $k_3=-0.5$, $\xi_{10}=-2.9$, $\xi_{20}=-5.9$, $\xi_{30}=0$ and $d=-1.1$.}
\end{figure}

{\textbf{\emph{Remark 2.}}} Due to the multiloop soliton solutions obtained in this section, the corresponding multibreather solutions and soliton molecules would be constructed directly.
Under certain condition imposed on the parameters, the breather solution is shown to
yield a nonsingular oscillating pulse solution of the WKI-SP$^{(1,1)}$
equation, which we shall term the breather solution as well. In fact, the $M$-breather solution of the  WKI-SP$^{(1,1)}$ equation can be
constructed from the $M$-breather solution of the MKdV-SG$^{(1,1)}$
equation \eqref{4-39}-\eqref{4-41} with $N=2M$ by choosing the
parameters appropriately. So we specify the parameters as
\begin{eqnarray}
k_{2j-1}=k_{2j}^{*},\quad \xi_{2j-1,0}=\xi_{2j,0}^{*} \label{BB1}
\end{eqnarray}
to satisfy
\begin{eqnarray}
\xi_{2j-1}=\xi_{2j}^{*}. \label{BB2}
\end{eqnarray}
The parametric solution \eqref{4-42-u} and \eqref{4-42-x} with \eqref{BB1} and \eqref{BB2}
describes multiple collisions of $M$ breathers.

Soliton molecules \cite{molecule1,molecule2,molecule3,molecule4} can be formed in some possible mechanisms both theoretically and experimentally. Here, we can obtain loop soliton molecules of the WKI-SP$^{(1,1)}$ equation by introducing the velocity resonance conditions\cite{Lou1}
\begin{eqnarray}
\frac{k_i}{k_j}=\frac{\alpha k_i^3-\frac{\beta}{k_i}}{\alpha k_j^3-\frac{\beta}{k_j}}.\label{velocity-resonance}
\end{eqnarray}
Two loop solitons are bounded to form a loop soliton molecule which are shown in
Fig.9(a). One can see that the distance between two loops will not be changed
as time goes on. Further, one can also immediately construct the  breather molecules by using the velocity resonant mechanism. Two breathers constitute one breather molecule which is shown in Fig.9(b). It is worth mentioning that such loop-type soliton molecules and
nonsingular oscillating pulse type breather molecules are firstly reported theoretically.



\begin{figure}[H]
	\centering
		\subfigure[]{
		\begin{minipage}[t]{0.48\textwidth}
		\centering
		\includegraphics[scale=0.5]{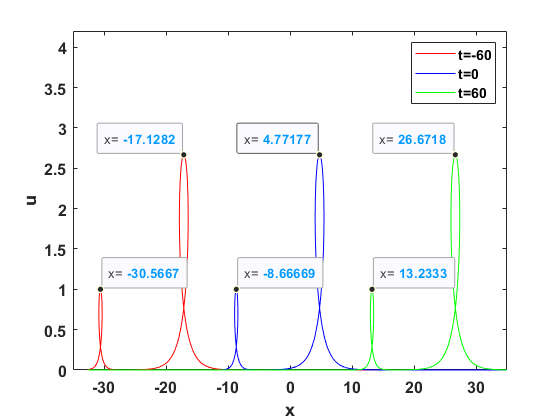}%
		 \end{minipage}%
		}%
		\subfigure[]{
		\begin{minipage}[t]{0.48\textwidth}
		\centering
		\includegraphics[scale=0.5]{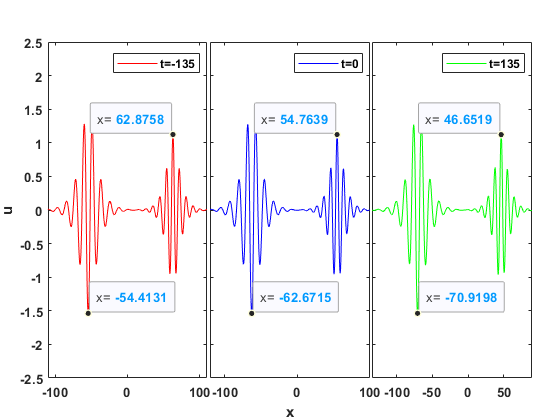}%
		\end{minipage}%
		}%
		\centering
		\caption{(a) A loop-type soliton molecule (b) a breather molecule}
\end{figure}

\section{Conclusion}

In this paper, we have constructed the generalized compound WKI-SP$^{(n,m)}$ equations
by virtue of two compatible linear spectral equations. By solving two sets of recursive
equations, we showed three SP-type equations (named WKI-SP$^{(0,1)}$, WKI-SP$^{(0,2)}$ and WKI-SP$^{(0,3)}$), two WKI-type equations (named WKI-SP$^{(1,0)}$ and WKI-SP$^{(2,0)}$) and
three compound WKI-SP equations (called WKI-SP$^{(1,1)}$, WKI-SP$^{(1,2)}$ and WKI-SP$^{(2,1)}$).
The classical WKI equation \eqref{1-1} is just the WKI-SP$^{(1,0)}$ equation while the
the SP equation equation \eqref{1-2} is the WKI-SP$^{(0,1)}$  equation. The WKI-SP$^{(0,2)}$ and WKI-SP$^{(0,3)}$ equations are the high-order short pulse equations.

It is known that there is a novel hodograph between the SP equation \eqref{1-2}
and the SG equation. And there also exits the hodograph transformation
between the WKI equation \eqref{1-1} and the (potential) MKdV equation. In this paper, with the help of one conservation law, we have found one hodograph transformation
which successfully transformed the compound WKI-SP$^{(1,1)}$ equation into the
compound MKdV-SG$^{(1,1)}$ equation. Meanwhile, the same hodograph transformation
could also enable us to convert  the compound WKI-SP$^{(1,2)}$ equation and the WKI-SP$^{(2,1)}$ equation into the MKdV-SG$^{(1,2)}$ equation and the MKdV-SG$^{(2,1)}$ equation respectively.

We constructed multiloop soliton solutions for the WKI-SP$^{(1,1)}$ equation as applications. The loop (antiloop) soliton solutions arise from the kink
(antikink) solutions of the MKdV-SG$^{(1,1)}$ equation. Let $m$ and $N-m$ be
the number of positive and negative $k_j$, respectively. Then,
the corresponding soliton solution would include $m$ loop
solitons and $N-m$ antiloop solitons. Here, we also call antiloop
as loop for convenience.
The analytic solutions for the one-loop, two-loop and three-loop solitons in the  parametric forms were shown. Especially, we stressed the description of two-loop soliton, which exhibit abundant solitonic behaviors.  On the one hand, two loop solitons
can either move in the same direction or in the opposite direction of the $x$-axis.
It is worth mentioned that there could be a small loop  with a fast speed
chasing a large loop  with a slow speed. Certainly, we also obtained
the typical solitonic behavior that the large loop  moves more rapidly than the
small loop.  On the other hand, there are two kinds of collision processes
depending  on the ratio of the eigenvalues involved. In the first case of two loop solitons
with dissimilar amplitudes chasing each other, the  small one would  travel around the
larger one. The phase shift of the small loop soliton can be seen to originate mainly from the delay caused by its travelling around the large loop soliton. In the second case of
the collision process for two loop solitons with similar amplitudes,
the slow loop soliton is pushed out forward with a considerable phase shift
when compared with that of the fast loop. That is to say the slow loop soliton
pass through the fast one.

It is also remarkable that all the models given by the WKI-SP$^{(n,m)}$
hierarchy here could be with time-varying coefficients. Hence, their dispersion relations will
have a time dependent velocity and the solitons will accelerate. These equations
with variable coefficients may be nice candidates in applications having accelerated ultra-short optical pulses. The method and the results mentioned in this paper would be extended to
the complex WKI-SP hierarchy, the nonlocal WKI-SP hierarchy, the semi-discrete and the discrete
WKI-SP hierarchy, the multi-component WKI-SP hierarchy (and with their corresponding complex, nonlocal, semi-discrete and discrete forms). These problems are to be pursued in our near future
work. The Darboux transformation, B\"{a}cklund transformation and  the Riemann-Hilbert problem
for the compound  WKI-SP equation  are also in our near consideration.

\section*{Acknowledgments}

This work is supported by the national natural science foundation of China (Grant No.
11771395 and No. 11871336) and the Zhejiang Provincial Natural Science Foundation of China (Grant
No. LY18A010034).

 \end{document}